\documentclass[journal]{IEEEtran}

\usepackage[dvips]{color}
\usepackage{graphicx}
\usepackage{amsmath}

\begin{document}

\title{Relay-Selection Improves the Security-Reliability Trade-off in Cognitive Radio Systems} \normalsize

%
%
%

\markboth{IEEE Transactions on Communications (accepted to appear)}%
{Yulong Zou \MakeLowercase{\textit{et al.}}: Relay-Selection Improves the Security-Reliability Trade-off in Cognitive Radio Systems}

\author{Yulong~Zou,~\IEEEmembership{Senior Member,~IEEE,}
        Benoit~Champagne,~\IEEEmembership{Senior Member,~IEEE,}
        Wei-Ping~Zhu,~\IEEEmembership{Senior Member,~IEEE,} and
        Lajos~Hanzo,~\IEEEmembership{Fellow,~IEEE}

\thanks{Manuscript received May 7, 2014; revised August 15, 2014 and October 16, 2014; accepted November 27, 2014. The editor coordinating the review of this paper and approving it for publication was Prof. Husheng Li.}
\thanks{This work was partially supported by the Priority Academic Program Development of Jiangsu Higher Education Institutions, the National Natural Science Foundation of China (Grant Nos. 61302104 and 61401223), the Scientific Research Foundation of Nanjing University of Posts and Telecommunications (Grant Nos. NY213014 and NY214001), the 1311 Talent Program of Nanjing University of Posts and Telecommunications, the Natural Science Foundation of Jiangsu Province (Grant No. BK20140887), and the Programme de bourses d'excellence pour etudiants etrangers (PBEEE) of the Government of Quebec.}
\thanks{Y. Zou is with the School of Telecommunications and Information Engineering, Nanjing University of Posts and Telecommunications, Nanjing, P. R. China. (Email: yulong.zou@njupt.edu.cn)}
\thanks{B. Champagne is with the Department of Electrical \& Computer Engineering, McGill University, Montreal, Canada. (Email: benoit.champagne@mcgill.ca)}
\thanks{W.-P. Zhu is with the Department of Electrical \& Computer Engineering, Concordia University, Montreal, Canada. (Email: weiping@ece.concordia.ca)}
\thanks{L. Hanzo is with the Department of Electronics and Computer Science, University of Southampton, Southampton, United Kingdom. (E-mail: lh@ecs.soton.ac.uk)}
}

\maketitle

\begin{abstract}
We consider a cognitive radio (CR) network consisting of a secondary transmitter (ST), a secondary destination (SD) and multiple secondary relays (SRs) in the presence of an eavesdropper, where the ST transmits to the SD with the assistance of SRs, while the eavesdropper attempts to intercept the secondary transmission. We rely on careful relay selection for protecting the ST-SD transmission against the eavesdropper with the aid of both single-relay and multi-relay selection. To be specific, only the ``best" SR is chosen in the single-relay selection for assisting the secondary transmission, whereas the multi-relay selection invokes multiple SRs for simultaneously forwarding the ST's transmission to the SD. We analyze both the intercept probability and outage probability of the proposed single-relay and multi-relay selection schemes for the secondary transmission relying on realistic spectrum sensing. We also evaluate the performance of classic direct transmission and artificial noise based methods for the purpose of comparison with the proposed relay selection schemes. It is shown that as the intercept probability requirement is relaxed, the outage performance of the direct transmission, the artificial noise based and the relay selection schemes improves, and vice versa. This implies a trade-off between the security and reliability of the secondary transmission in the presence of eavesdropping attacks, which is referred to as the \emph{security-reliability trade-off} (SRT). Furthermore, we demonstrate that the SRTs of the single-relay and multi-relay selection schemes are generally better than that of classic direct transmission, explicitly demonstrating the advantage of the proposed relay selection in terms of protecting the secondary transmissions against eavesdropping attacks. Moreover, as the number of SRs increases, the SRTs of the proposed single-relay and multi-relay selection approaches significantly improve. Finally, our numerical results show that as expected, the multi-relay selection scheme achieves a better SRT performance than the single-relay selection.

\end{abstract}


\section{Introduction}

\IEEEPARstart {T}{he} security aspects of cognitive radio (CR) systems [1]-[3] have attracted increasing attention from the research community. Indeed, due to the highly dynamic nature of the CR network architecture, legitimate CR devices become exposed to both internal as well as to external attackers and hence they are extremely vulnerable to malicious behavior. For example, an illegitimate user may intentionally impose interference (i.e. jamming) for the sake of artificially contaminating the CR environment [4]. Hence, the CR users fail to accurately characterize their surrounding radio environment and may become misled or compromised, which leads to a malfunction. Alternatively, an illegitimate user may attempt to tap the communications of authorized CR users by eavesdropping, in order to intercept confidential information.

Clearly, CR networks face diverse security threats during both spectrum sensing [5], [6] as well as spectrum sharing [7], spectrum mobility [8] and spectrum management [9]. Extensive studies have been carried out for protecting CR networks both against primary user emulation (PUE) [10] and against denial-of-service (DoS) attacks [11]. In addition to PUE and DoS attacks, eavesdropping is another main concern in protecting the data confidentiality [12], although it has received less attention in the literature on CR network security. Traditionally, cryptographic techniques are employed for guaranteeing transmission confidentiality against an eavesdropping attack. However, this introduces a significant computational overhead [13] as well as imposing additional system complexity in terms of the secret key management [14]. Furthermore, the existing cryptographic approaches are not perfectly secure and can still be decrypted by an eavesdropper (E), provided that it has the capacity to carry out exhaustive key search with the aid of brute-force attack [15].

Physical-layer security [16], [17] is emerging as an efficient approach for defending authorized users against eavesdropping attacks by exploiting the physical characteristics of wireless channels. In [17], Leung-Yan-Cheong and Hellman demonstrated that perfectly secure and reliable transmission can be achieved, when the wiretap channel spanning from the source to the eavesdropper is a further degraded version of the main channel between the source and destination. They also showed that the maximal secrecy rate achieved at the legitimate destination, which is termed the secrecy capacity, is the difference between the capacity of the main channel and that of the wiretap channel. In [18]-[20], the secrecy capacity limits of wireless fading channels were further developed and characterized from an information-theoretic perspective, demonstrating the detrimental impact of wireless fading on the physical-layer security. In order to combat the fading effects, both multiple-input multiple-output (MIMO) schemes [21], [22] as well as cooperative relaying [23]-[25] and beamforming techniques [26], [27] were investigated for the sake of enhancing the achievable wireless secrecy capacity. Although extensive research efforts were devoted to improving the security of traditional wireless networks [16]-[27], less attention has been dedicated to CR networks. In [28] and [29], the achievable secrecy rate of the secondary transmission was investigated under a specific quality-of-service (QoS) constraint imposed on the primary transmission. {Additionally, an overview of the physical-layer security aspects of CR networks was provided in [30], where several security attacks as well as the related countermeasures are discussed.} In contrast to conventional non-cognitive wireless networks, the physical-layer security of CR networks has to consider diverse additional challenges, including the protection of the primary user's QoS and the mitigation of the mutual interference between the primary and secondary transmissions.

Motivated by the above considerations, we explore the physical-layer security of a CR network comprised of a secondary transmitter (ST) communicating with a secondary destination (SD) with the aid of multiple secondary relays (SRs) in the presence of an unauthorized attacker. Our main focus is on investigating the security-reliability trade-off (SRT) of the cognitive relay transmission in the presence of realistic spectrum sensing. The notion of the SRT in wireless physical-layer security was introduced and examined in [31], where the security and reliability was characterized in terms of the intercept probability and outage probability, respectively. In contrast to the conventional non-cognitive wireless networks studied in [31], the SRT analysis of CR networks presented in this work additionally takes into account the mutual interference between the primary user (PU) and secondary user (SU).

{The main contributions of this paper are summarized as follows.}
\begin{itemize}
\item
We propose two relay selection schemes, namely both single-relay and multi-relay selection, for protecting the secondary transmissions against eavesdropping attacks. More specifically, in the single-relay selection (SRS) scheme, only a single relay is chosen from the set of multiple SRs for forwarding the secondary transmissions from the ST to the SD. By contrast, the multi-relay selection (MRS) scheme employs multiple SRs for simultaneously assisting the ST-SD transmissions.
\item
We present the mathematical SRT analysis of the proposed SRS and MRS schemes in the presence of realistic spectrum sensing. Closed-form expressions are derived for the intercept probability (IP) and outage probability (OP) of both schemes for transmission over Rayleigh fading channels. The numerical SRT results of conventional direct transmission and {artificial noise based} schemes are also provided for comparison purposes.
\item
It is shown that as the spectrum sensing reliability is increased and/or the false alarm probability is reduced, the SRTs of both the SRS and MRS schemes are improved. Numerical results demonstrate that the proposed SRS and MRS schemes generally outperform the conventional direct transmission and {artificial noise based} approaches in terms of their SRTs.
\end{itemize}

The remainder of this paper is organized as follows. Section II presents the system model of physical-layer security in CR networks in the context of both the direct transmission as well as the SRS and MRS schemes. In Section III, we analyze the SRTs of these schemes in the presence of realistic spectrum sensing over Rayleigh fading channels. Next, numerical SRT results of the direct transmission, SRS and MRS schemes are given in Section IV, {where the SRT performance of the artificial noise based scheme is also numerically evaluated for comparison purposes.} Finally, Section V provides our concluding remarks.

\section{Relay Selection Aided Protection against Eavesdropping in CR Networks}
We first introduce the overall system model of physical-layer security in CR networks. We then present the signal model of the conventional direct transmission approach, which will serve as our benchmarker, as well as of the SRS and MRS schemes for improving the CR system's security against eavesdropping attacks.
\subsection{System Model}
\begin{figure}
  \centering
  {\includegraphics[scale=0.6]{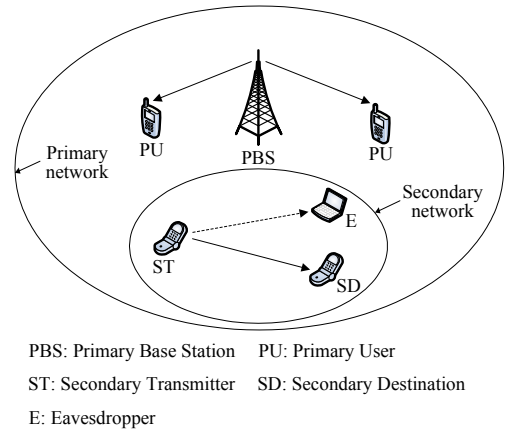}\\
  \caption{A primary wireless network in coexistence with a secondary CR network.}\label{Fig1}}
\end{figure}

As shown in Fig. 1, we consider a primary network in coexistence with a secondary network (also referred to as a \emph{CR network}). The primary network includes a primary base station (PBS) and multiple primary users (PUs), which communicate with the PBS over the licensed spectrum. By contrast, the secondary network consisting of one or more STs and SDs exploits the licensed spectrum in an opportunistic way. To be specific, a particular ST should first detect with the aid of spectrum sensing whether or not the licensed spectrum is occupied by the PBS. If so, the ST is not at liberty to transmit in order to avoid interfering with the PUs. If alternatively, the licensed spectrum is deemed to be unoccupied (i.e. a spectrum hole is detected), then the ST may transmit to the SD over the detected spectrum hole. Meanwhile, E attempts to intercept the secondary transmission from the ST to the SD. {For notational convenience, let ${H_0}$ and $H_1$ represent the event that the licensed spectrum is unoccupied and occupied by the PBS during a particular time slot, respectively.} Moreover, let ${\hat H}$ denote the status of the licensed spectrum detected by spectrum sensing. Specifically, ${\hat H} = {H_0}$ represents the case that the licensed spectrum is deemed to be unoccupied, while ${\hat H} = {H_1}$ indicates that the licensed spectrum is deemed to be occupied.

The probability $P_d$ of correct detection of the presence of PBS and the associated false alarm probability $P_f$ are defined as $P_d  = \Pr ( {\hat H  = H_1 |H_1 } ) $ and  $P_f  = \Pr ( {\hat H  = H_1 |H_0 } ) $, respectively. Due to the background noise and fading effects, it is impossible to achieve perfectly reliable spectrum sensing without missing the detection of an active PU and without false alarm, which suggests that a spectral band is occupied by a PU, when it is actually unoccupied. Moreover, the missed detection of the presence of PBS will result in interference between the PU and SU. {In order to guarantee that the interference imposed on the PUs is below a tolerable level, both the successful detection probability (SDP) ${{{P}}_{{d}}}$ and false alarm probability (FAP) ${{{P}}_{{f}}}$ should be within a meaningful target range. For example, the IEEE 802.22 standard requires $P_d>0.9$ and $P_f<0.1$ [2]. For better protection of PUs, we consider ${{{P}}_{{d}}} = 0.99$ and ${{{P}}_{{f}}} = 0.01$, unless otherwise stated.} Additionally, we consider a Rayleigh fading model for characterizing all the channels between any two nodes of Fig. 1. Finally, all the received signals are assumed to be corrupted by additive white Gaussian noise (AWGN) having a zero mean and a variance of $N_0$.

\subsection{Direct Transmission}
Let us first consider the conventional direct transmission as a benchmark scheme. Let $x_p$ and $x_s$ denote the random symbols transmitted by the PBS and the ST at a particular time instance. Without loss of generality, we assume $E[|x_p|^2]=E[|x_s|^2]=1$, {where $E[\cdot]$ represents the expected value operator}. The transmit powers of the PBS and ST are denoted by $P_p$ and $P_s $, respectively. Given that the licensed spectrum is deemed to be unoccupied by the PBS (i.e. ${\hat H} = {H_0}$), ST transmits its signal ${x_s}$ at a power of $P_s$. Then, the signal received at the SD can be written as
\begin{equation}\label{equa1}
y_d  = h_{sd} \sqrt {P_s } x_s  + h_{pd} \sqrt {\alpha P_p } x_p  + n_d,
\end{equation}
where $h_{sd}$ and $h_{pd}$ represent the fading coefficients of the channel spanning from ST to SD and that from PBS to SD, respectively. Furthermore, $n_d$ represents the AWGN received at SD and the random variable (RV) $\alpha$ is defined as
\begin{equation}\label{equa2}
\alpha = \left\{ \begin{array}{l}
 0,\quad H_0 \\
 1,\quad H_1, \\
 \end{array} \right.
\end{equation}
{where $H_0$ represents that the licensed spectrum is unoccupied by PBS and no primary signal is transmitted, leading to $\alpha=0$. By contrast, ${H_1}$ represents that PBS is transmitting its signal $x_p$ over the licensed spectrum, thus $\alpha=1$.} Meanwhile, due to the broadcast nature of the wireless medium, the ST's signal will be overheard by E and the overheard signal can be expressed as
\begin{equation}\label{equa3}
y_e  = h_{se} \sqrt {P_s } x_s  + h_{pe} \sqrt {\alpha P_p } x_p  + n_e,
\end{equation}
where $h_{se}$ and $h_{pe}$ represent the fading coefficients of the channel spanning from ST to E and that from PBS to E, respectively, while $n_e$ represents the AWGN received at E. Upon combining Shannon's capacity formula [31] with (1), we obtain the capacity of the ST-SD channel as
\begin{equation}\label{equa4}
C_{sd}  = \log _2 ( {1 + \frac{{|h_{sd} |^2 \gamma _s }}{{\alpha |h_{pd} |^2 \gamma _p  + 1}}} ),
\end{equation}
where $\gamma _s  = {{P_s }}/{{N_0 }}$ and $\gamma _p  = {{P_p }}/{{N_0 }}$. Similarly, the capacity of the ST-E channel is obtained from (3) as
\begin{equation}\label{equa5}
C_{se}  = \log _2 ( {1 + \frac{{|h_{se} |^2 \gamma _s }}{{\alpha |h_{pe} |^2 \gamma _p  + 1}}} ).
\end{equation}

\subsection{Single-Relay Selection}
\begin{figure}
  \centering
  {\includegraphics[scale=0.6]{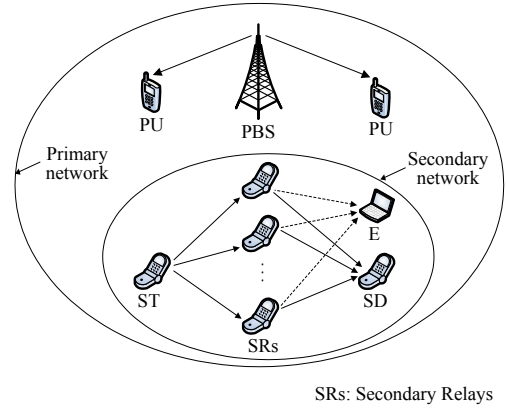}\\
  \caption{A cognitive relay network consists of one ST, one SD and $N$ SRs in the presence of an E.}\label{Fig2}}
\end{figure}

In this subsection, we consider the cognitive relay network of Fig. 2, {where both SD and E are assumed to be beyond the coverage area of the ST [24], [25]}, and $N$ secondary relays (SRs) are employed for assisting the cognitive ST-SD transmission. We assume that {a common control channel (CCC) [6] is available for coordinating the actions of the different network nodes} and the decode-and-forward (DF) relaying using two adjacent time slots is employed. More specifically, once the licensed spectrum is deemed to be unoccupied, the ST first broadcasts its signal $x_s$ to the $N$ SRs, which attempt to decode $x_s$ from their received signals. For notational convenience, let ${\cal {D}}$ represent the set of SRs that succeed in decoding $x_s$. Given $N$ SRs, there are $2^N$ possible subsets ${\cal {D}}$, thus the sample space of ${\cal {D}}$ is formulated as
\begin{equation}\label{equa6}
\Omega  = \left\{ {\emptyset ,{\cal {D}}_1 ,{\cal {D}}_2 , \cdots ,{\cal {D}}_n , \cdots ,{\cal {D}}_{2^N  - 1} } \right\},
\end{equation}
where $ \emptyset$ represents the empty set and $ {\cal {D}}_n$ represents the $n{\textrm{-th}}$ non-empty subset of the $N$ SRs. If the set $\cal {D}$ is empty, implying that no SR decodes $x_s$ successfully, then all the SRs remain silent and thus both SD and E are unable to decode $x_s$ in this case. If the set ${\cal {D}}$ is non-empty, a specific SR is chosen from ${\cal {D}}$ to forward its decoded signal $x_s$ to SD. Therefore, given ${\hat H} = {H_0}$ (i.e. the licensed spectrum is deemed unoccupied), ST broadcasts its signal ${x_s}$ to $N$ SRs at a power of $P_s$ and a rate of $R$. Hence, the signal received at a specific SR$_i$ is given by
\begin{equation}\label{equa7}
y_i  = h_{si} \sqrt {P_s } x_s  + h_{pi} \sqrt {\alpha P_p } x_p  + n_i,
\end{equation}
where $h_{si}$ and $h_{pi}$ represent the fading coefficients of the ST-SR$_i$ channel and that of the PBS-SR$_i$ channel, respectively, with $n_i$ representing the AWGN at SR$_i$. From (7), we obtain the capacity of the ST-SR$_i$ channel as
\begin{equation}\label{equa8}
C_{si}  = \frac{1}{2}\log _2 ( {1 + \frac{{|h_{si} |^2 \gamma _s }}{{\alpha |h_{pi} |^2 \gamma _p  + 1}}} ),
\end{equation}
where the factor $\frac{1}{2}$ arises from the fact that two orthogonal time slots are required for completing the message transmission from ST to SD via SR$_i$. According to Shannon's coding theorem, if the data rate is higher than the channel capacity, the receiver becomes unable to successfully decode the source signal, regardless of the decoding algorithm adopted. Otherwise, the receiver can succeed in decoding the source signal. Thus, using (8), we can describe the event of ${\cal {D}}=\emptyset$ as
\begin{equation}\label{equa9}
C_{si}  < R,\quad i \in \{1,2,\cdots,N\}.
\end{equation}
Meanwhile, the event of ${\cal {D}}={\cal {D}}_n$ is described as
\begin{equation}\label{equa10}
\begin{split}
&C_{si}  > R,\quad i \in {\cal {D}}_n  \\
&C_{sj}  < R,\quad j \in \bar {\cal {D}}_n,  \\
 \end{split}
\end{equation}
where $\bar {\cal {D}}_n$ represents the complementary set of ${\cal {D}}_n$. Without loss of generality, we assume that SR$_i$ is chosen within ${\cal {D}}_n$ to transmit its decoded result $x_s$ at a power of $P_s$, thus the signal received at SD can be written as
\begin{equation}\label{equa11}
y_d = h_{id} \sqrt {P_s } x_s  + h_{pd} \sqrt {\alpha P_p } x_p  + n_d,
\end{equation}
where $h_{id}$ represents the fading coefficient of the SR$_i$-SD channel. From (11), the capacity of the SR${_i}$-SD channel is given by
\begin{equation}\label{equa12}
C_{id} = \frac{1}{2}\log _2 ( {1 + \frac{{|h_{id} |^2 \gamma _s }}{{\alpha |h_{pd} |^2 \gamma _p  + 1}}} ),
\end{equation}
where $i \in {\cal{D}}_n$. In general, the specific SR${_i}$ having the highest instantaneous capacity to SD is chosen as the ``best" SR for assisting the ST's transmission. Therefore, the best relay selection criterion is expressed from (12) as
\begin{equation}\label{equa13}
{\textrm{Best SR}} = \arg \mathop {\max }\limits_{i \in {\cal {D}}_n } C_{id}  = \arg \mathop {\max }\limits_{i \in {\cal {D}}_n } |h_{id} |^2,
\end{equation}
which shows that only {the channel state information (CSI)} $|h_{id}|^2$ is required for performing the relay selection without the need for the eavesdropper's CSI knowledge. Upon combining (12) and (13), we obtain the capacity of the channel spanning from the ``best" SR to SD as
\begin{equation}\label{equa14}
C_{bd} = \frac{1}{2} \log _2 ( {1 + \frac{{ \gamma _s }}{{\alpha |h_{pd} |^2 \gamma _p  + 1}}\mathop {\max }\limits_{i \in {\cal{D}}_n } |h_{id} |^2 } ),
\end{equation}
where the subscript `$b$' in $C_{bd}$ denotes the best SR. {It is observed from (14) that the legitimate transmission capacity of the SRS scheme is determined by the maximum of independent random variables (RVs) $|h_{id}|^2$ for different SRs. By contrast, one can see from (4) that the capacity of classic direct transmission is affected by the single RV $|h_{sd}|^2$. If all RVs $|h_{id}|^2$ and $|h_{sd}|^2$ are independent and identically distributed (i.i.d),
it would be most likely that $\mathop {\max }\limits_{i \in {\cal{D}}_n } |h_{id} |^2$ is much higher than $|h_{sd}|^2$ for a sufficiently large number of SRs, resulting in a performance improvement for the SRS scheme over the classic direct transmission. However, if the RVs $|h_{id}|^2$ and $|h_{sd}|^2$ are non-identically distributed and the mean value of $|h_{sd}|^2$ is much higher than that of $|h_{id}|^2$, then it may be more likely that $\mathop {\max }\limits_{i \in {\cal{D}}_n } |h_{id} |^2$ is smaller than $|h_{sd}|^2$ for a given number of SRs. In this extreme case, the classic direct transmission may perform better than the SRS scheme. It is worth mentioning that in practice, the average fading gain of the SR$_i$-SD channel, $|h_{id}|^2$, should not be less than that of the ST-SD channel $|h_{sd}|^2$, since SRs are typically placed in the middle between the ST and SD. Hence, a performance improvement for the SRS scheme over classic direct transmission would be achieved in practical wireless systems. Note that although a factor $\frac{1}{2}$ in (14) is imposed on the capacity of the main channel, it would not affect the performance of the SRS scheme from a SRT perspective, since the capacity of the wiretap channel is also multiplied by $\frac{1}{2}$ as will be shown in (16).}

Additionally, given that the selected SR transmits its decoded result $x_s$ at a power of $P_s$, the signal received at E is expressed as
\begin{equation}\label{equa15}
y_e  = h_{be} \sqrt {P_s } x_s  + h_{pe} \sqrt {\alpha P_p } x_p  + n_e,
\end{equation}
where $h_{be}$ and $h_{pe}$ represent the fading coefficients of the channel from ``best" SR to E and that from PBS to E, respectively. From (15), the capacity of the channel spanning from the ``best" SR to E is given by
\begin{equation}\label{equa16}
C_{be} = \frac{1}{2}\log _2 ( {1 + \frac{{|h_{be} |^2 \gamma _s }}{{\alpha |h_{pe} |^2 \gamma _p  + 1}}} ),
\end{equation}
where $b \in {\cal{D}}_n$ is determined by the relay selection criterion given in (13). {{As shown in (16), the eavesdropper's channel capacity is affected by the channel state information (CSI) $|h_{be}|^2$ of the wiretap channel spanning from the ``best" relay to the eavesdropper. However, one can see from (13) that the best relay is selected from the decoding set ${\cal D}_n$ solely based on the main channel's CSI $|{h_{id}}{|^2}$ i.e. without taking into account the eavesdropper's CSI knowledge of $|{h_{ie}}{|^2}$. This means that the selection of the best relay aiming for maximizing the legitimate transmission capacity of (14) would not lead to significantly beneficial or adverse impact on the eavesdropper's channel capacity, since the main channel and the wiretap channel are independent of each other.}}

{{For example, if the random variables (RVs) $|h_{ie}|^2$ related to the different relays are i.i.d, we can readily infer by the law of total probability that $|h_{be}|^2$ has the same probability density function (PDF) as $|h_{ie}|^2$, implying that the eavesdropper's channel capacity of (16) is not affected by the selection of the best relay given by (13). Therefore, the SRS scheme has no obvious advantage over the classic direct transmission in terms of minimizing the capacity of the wiretap channel. To elaborate a little further, according to the SRT trade-off, a reduction of the outage probability (OP) due to the capacity enhancement of the main channel achieved by using the selection of the best relay would be converted into an intercept probability (IP) improvement, which will be numerically illustrated in Section IV.}}

\subsection{Multi-Relay Selection}
This subsection presents a MRS scheme, where multiple SRs are employed for simultaneously forwarding the source signal $x_s$ to SD. To be specific, ST first transmits $x_s$ to $N$ SRs over a detected spectrum hole. As mentioned in Subsection II-C, we denote by ${\cal {D}}$ the set of SRs that successfully decode $x_s$. If $\cal {D}$ is empty, all SRs fail to decode $x_s$ and will not forward the source signal, thus both SD and E are unable to decode $x_s$. If ${\cal {D}}$ is non-empty (i.e. ${\cal {D}}={\cal {D}}_n$), all SRs within ${\cal {D}}_n$ are utilized for simultaneously transmitting $x_s$ to SD. This differs from the SRS scheme, where only a single SR is chosen from ${\cal {D}}_n$ for forwarding $x_s$ to SD. In order to make effective use of multiple SRs, a weight vector denoted by ${{{w}}} = [w_1 ,w_2 , \cdots ,w_{|{\cal{D}}_n|} ]^T$ is employed at the SRs for transmitting $x_s$, where $|{\cal {D}}_n|$ is the cardinality of the set ${\cal{D}}_n$. For the sake of a fair comparison with the SRS scheme in terms of power consumption, the total transmit power across all SRs within ${\cal {D}}_n$ shall be constrained to $P_s$ and thus the weight vector ${{{w}}}$ should be normalized according to $||{{w}}|| = 1$. Thus, given ${\cal {D}}={\cal {D}}_n$ and considering that all SRs within ${\cal {D}}_n$ are selected for simultaneously transmitting $x_s$ with a weight vector ${{{w}}}$, the signal received at SD is expressed as
\begin{equation}\label{equa17}
y_d^{{\textrm{multi}}}  = \sqrt {{P_s }} {{w}}^T {{H}}_d x_s  + \sqrt {\alpha P_p } h_{pd} x_p  + n_d,
\end{equation}
where ${{H}}_d  = [h_{1d} ,h_{2d} , \cdots ,h_{|{\cal{D}}_n |d} ]^T$. Similarly, the signal received at E can be written as
\begin{equation}\label{equa18}
y_e^{{\textrm{multi}}}  = \sqrt {{P_s }} {{w}}^T {{H}}_e x_s  + \sqrt {\alpha P_p } h_{pe} x_p  + n_e,
\end{equation}
where ${{H}}_e  = [h_{1e} ,h_{2e} , \cdots ,h_{|{\cal{D}}_n |e} ]^T$. From (17) and (18), the signal-to-interference-plus-noise ratios (SINRs) at SD and E are, respectively, given by
\begin{equation}\label{equa19}
{\textrm{SINR}}^{{\textrm{multi}}}_d  = \frac{{\gamma _s }}{{\alpha |h_{pd} |^2 \gamma _p + 1}}|{{w}}^T {{H}}_d |^2,
\end{equation}
and
\begin{equation}\label{equa20}
{\textrm{SINR}}^{{\textrm{multi}}}_e  = \frac{{\gamma _s }}{{\alpha |h_{pe} |^2 \gamma _p  + 1}}|{{w}}^T {{H}}_e |^2.
\end{equation}
In this work, the weight vector ${{{w}}}$ is optimized by maximizing the SINR at SD, yielding
\begin{equation}\label{equa21}
\begin{split}
\mathop {\max }\limits_{{w}} {\textrm{SINR}}^{{\textrm{multi}}}_d ,\quad {\textrm{s.t. }}||{{w}}|| = 1,
\end{split}
\end{equation}
where the constraint is used for normalization purposes. Using the Cauchy-Schwarz inequality [32], we can readily obtain the optimal weight vector ${{w}}_{{\textrm{opt}}}$ from (21) as
\begin{equation}\label{equa22}
{{w}}_{{\textrm{opt}}}  = \frac{{{{H}}_d^* }}{{|{{H}}_d |}},
\end{equation}
which indicates that the optimal vector design only requires the SR-SD CSI ${{H}}_d$, whilst dispensing with the eavesdropper's CSI ${{H}}_e$. Substituting the optimal vector ${{w}}_{{\textrm{opt}}}$ from (22) into (19) and (20) and using Shannon's capacity formula, we can obtain the channel capacities achieved at both SD and E as
\begin{equation}\label{equa23}
C_d^{{\textrm{multi}}}  = \frac{1}{2}\log _2 ( {1 + \frac{{\gamma _s }}{{\alpha \gamma _p |h_{pd} |^2  + 1}}\sum\limits_{i \in {\cal{D}}_n } {|h_{id} |^2 } } ),
\end{equation}
and
\begin{equation}\label{equa24}
C_e^{{\textrm{multi}}}  = \frac{1}{2}\log _2 ( {1 + \frac{{\gamma _s }}{{\alpha \gamma _p |h_{pe} |^2  + 1}}\frac{{|{{H}}_d^{H}{{H}}_e |^2 }}{{|{{H}}_d |^2 }}} ),
\end{equation}
for ${\cal{D}} = {\cal{D}}_n $, where $H$ represents the Hermitian transpose. {One can observe from (14) and (23) that the difference between the capacity expressions $C_{bd}$ and $C_d^{{\textrm{multi}}} $ only lies in the fact that the maximum of RVs $|h_{id} |^2$ for different SRs (i.e., $\mathop {\max }\limits_{i \in {\cal{D}}_n } |h_{id} |^2$) is used for the SRS scheme, while the sum of RVs $|h_{id} |^2$ (i.e., $\sum\limits_{i \in {\cal{D}}_n } {|h_{id} |^2 }$) is employed for the MRS scheme. Clearly, we have $\sum\limits_{i \in {\cal{D}}_n } {|h_{id} |^2 } > \mathop {\max }\limits_{i \in {\cal{D}}_n } |h_{id} |^2$, resulting in a performance gain for MRS over SRS in terms of maximizing the legitimate transmission capacity. {{Moreover, since the main channel ${{H}}_d$ and the wiretap channel ${{H}}_e$ are independent of each other, the optimal weights assigned for the multiple relays based on ${{H}}_d$ will only slightly affect the eavesdropper's channel capacity. This means that the MRS and SRS schemes achieve more or less the same performance in terms of the capacity of the wiretap channel. Nevertheless, given a fixed outage requirement, the MRS scheme can achieve a better intercept performance than the SRS scheme, because according to the SRT, an outage reduction achieved by the capacity enhancement of the legitimate transmission relying on the MRS would be converted into an intercept improvement. To be specific, given an enhanced capacity of the legitimate transmission, we may increase the data rate $R$ based on the OP definition of (25) for maintaining a fixed OP, which, in turn leads to a reduction of the IP, since a higher data rate would result in a lower IP, according to the IP definition of (26).}}

It needs to be pointed out that in the MRS scheme, a high-complexity symbol-level synchronization is required for multiple distributed SRs, when simultaneously transmitting to SD, whereas the SRS does not require such a complex synchronization process. Thus, the performance improvement of MRS over SRS is achieved at the cost of a higher implementation complexity. Additionally, the synchronization imperfections of the MRS scheme will impose a performance degradation, which may even lead to a performance for the MRS scheme becoming worse than that of the SRS scheme.

Throughout this paper, the Rayleigh model is used for characterizing the fading amplitudes (e.g., $|h_{sd}|$, $|h_{si}|$, $|h_{id}|$, etc.) of wireless channels, which, in turn, implies that the fading square magnitudes $|h_{sd}|^2$, $|h_{si}|^2$ and $|h_{id}|^2$ are exponentially distributed random variables (RVs).} So far, we have completed the presentation of the signal model of the direct transmission, of the SRS, and of the MRS schemes for CR networks applications in the presence of eavesdropping attacks.


\section{SRT Analysis over Rayleigh Fading Channels}
This section presents the SRT analysis of the direct transmission, SRS and MRS schemes over Rayleigh fading channels. As discussed in [31], the security and reliability are quantified in terms of the IP and OP experienced by the eavesdropper and destination, respectively. {It is pointed out that in CR networks, ST starts to transmit its signal only when an available spectrum hole is detected. Similarly to [34], the OP and IP are thus calculated under the condition that the licensed spectrum is detected to be unoccupied by the PBS. The following gives the definition of OP and IP}.\\
{Definition 1}: \emph{Let $C_{d}$ and $C_{e}$ represent the channel capacities achieved at the destination and eavesdropper, respectively. The OP and IP are, respectively, defined as}
\begin{equation}\label{equa25}
{P_{{\textrm{out}}} = \Pr ( {{C_{d}} < {R} | {\hat H  = H_0 }} ),}
\end{equation}
\emph{and}
\begin{equation}\label{equa26}
{P_{{\textrm{int}}} = \Pr ( {{C_{e}} > {R} | {\hat H  = H_0 }} ),}
\end{equation}
\emph{where $R$ is the data rate.}

\subsection{Direct Transmission}
Let us first analyze the SRT performance of the conventional direct transmission. Given that a spectrum hole has been detected, the OP of direct transmission is obtained from (25) as
\begin{equation}\label{equa27}
P_{{\textrm{out}}}^{{\textrm{direct}}}  = \Pr ( {C_{sd}  < R | {\hat H  = H_0 } } ),
\end{equation}
where $C_{{sd}}$ is given by (4). Using the law of total probability, we can rewrite (27) as
\begin{equation}\label{equa28}
P_{{\textrm{out}}}^{{\textrm{direct}}} = \Pr ( {C_{sd}  < R,H_0 | {\hat H  = H_0 } } ) + \Pr ( {C_{sd}  < R,H_1 | {\hat H  = H_0 } } ),
\end{equation}
which can be further expressed as
\begin{equation}\label{equa29}
\begin{split}
 P_{{\textrm{out}}}^{{\textrm{direct}}}  =& \Pr ( {C_{sd}  < R| {H_0, \hat H  = H_0 } } )\Pr ( {H_0 | {\hat H  = H_0 } } ) \\
  &+ \Pr ( {C_{sd}  < R| {H_1, \hat H  = H_0 } } )\Pr ( {H_1 | {\hat H  = H_0 } } ). \\
 \end{split}
\end{equation}
{It is shown from (2) that given $H_0$ and $H_1$, the parameter $\alpha$ is obtained as $\alpha=0$ and $\alpha=1$, respectively. Thus, combining (2) and (4), we have $C_{sd} = \log _2 (1 + |h_{sd} |^2 \gamma _s )$ given $ H_0 $ and $C_{sd} =  \log _2 (1 + \frac{{|h_{sd} |^2 \gamma _s }}{{|h_{pd} |^2 \gamma _p  + 1}})$ given $H_1$. Substituting this result into (29) yields}
\begin{equation}\label{equa30}
\begin{split}
 P_{{\textrm{out}}}^{{\textrm{direct}}}  =& \Pr ( {|h_{sd} |^2 \gamma _s  < 2^R  - 1} )\Pr ( {H_0 | {\hat H  = H_0 } } )\\
  &+ \Pr ( {\frac{{|h_{sd} |^2 \gamma _s }}{{|h_{pd} |^2 \gamma _p  + 1}} < 2^R  - 1} )\Pr ( {H_1 | {\hat H  = H_0 } } ).
 \end{split}
\end{equation}
Moreover, the terms $\Pr ( {H_0 | {\hat H  = H_0 } } )$ and $\Pr ( {H_1 | {\hat H  = H_0 } } )$ can be obtained by using Bayes' theorem as
\begin{equation}\label{equa31}
\begin{split}
 \Pr ( {H_0 | {\hat H  = H_0 } } ) = & \frac{{\Pr ( {\hat H  = H_0 | {H_0 } } )\Pr ( {H_0 } )}}{{\sum\limits_{i \in \{0,1\}} {\Pr ( {\hat H  = H_0 | {H_i } } )\Pr ( {H_i } )} }} \\
  = &\frac{{P_0 (1 - P_f )}}{{P_0 (1 - P_f ) + (1 - P_0 )(1 - P_d )}} \buildrel \Delta \over = \pi _0,
 \end{split}
\end{equation}
and
\begin{equation}\label{equa32}
\Pr ( {H_1 | {\hat H  = H_0 } } ) = \frac{{(1 - P_0 )(1 - P_d )}}{{P_0 (1 - P_f ) + (1 - P_0 )(1 - P_d )}}\buildrel \Delta \over = \pi _1,
\end{equation}
where $P_0  = \Pr ( {H_0 } )$ is the probability that the licensed spectrum band is unoccupied by PBS, while ${{{P}}_{{d}}} = \Pr ({\hat H} = {H_1}|{H_1})$ and ${{{P}}_{{f}}} = \Pr ({\hat H} = {H_1}|{H_0})$ are the SDP and FAP, respectively. For notational convenience, we introduce the shorthand $\pi _0  = \Pr ( {H_0 | {\hat H  = H_0 } } )$, $\pi _1  = \Pr ( {H_1 | {\hat H  = H_0 } } )$ and $\Delta  = \frac{{2^R  - 1}}{{\gamma _s }}$. Then, using (31) and (32), we rewrite (30) as
\begin{equation}\label{equa33}
P_{{\textrm{out}}}^{{\textrm{direct}}}  = \pi _0 \Pr ( {|h_{sd} |^2  < \Delta } ) + \pi _1 \Pr ( {|h_{sd} |^2  - |h_{pd} |^2 \gamma _p \Delta  < \Delta } ).
\end{equation}
Noting that $|{h_{sd}}{|^2}$ and $|{h_{pd}}{|^2}$ are independently and exponentially distributed RVs with respective means of $\sigma _{sd}^{{\rm{  }}2}$ and $\sigma _{pd}^{{\rm{  }}2}$, we obtain
\begin{equation}\label{equa34}
\Pr (|{h_{sd}}{|^2} < \Delta ) = 1 - \exp ( - \frac{\Delta }{{\sigma _{sd}^2}}),
\end{equation}
and
\begin{equation}\label{equa35}
\Pr (|{h_{sd}}{|^2} - |{h_{pd}}{|^2} {\gamma _p}\Delta < \Delta ) = 1 - \frac{{\sigma _{sd}^2}}{{\sigma _{pd}^2{\gamma _p}\Delta  + \sigma _{sd}^2}}\exp ( - \frac{\Delta }{{\sigma _{sd}^2}}).
\end{equation}

Additionally, we observe from (26) that an intercept event occurs, when the capacity of the ST-E channel becomes higher than the data rate. Thus, given that a spectrum hole has been detected (i.e. $\hat H  = H_0 $), ST starts transmitting its signal to SD and E may overhear the ST-SD transmission. The corresponding IP is given by
\begin{equation}\label{equa36}
P_{{\textrm{int}}}^{{\textrm{direct}}}  = \Pr ( {C_{se}  > R| {\hat H  = H_0 } } ),
\end{equation}
which can be further expressed as
\begin{equation}\label{equa37}
\begin{split}
 P_{{\textrm{int}}}^{{\textrm{direct}}}  = &\Pr ( {C_{se}  > R| {\hat H  = H_0 ,H_0 } } )\Pr ( {H_0 | {\hat H  = H_0 } } ) \\
  &+ \Pr ( {C_{se}  > R| {\hat H  = H_0 ,H_1 } } )\Pr ( {H_1 | {\hat H  = H_0 } } ) \\
  = &\pi _0 \Pr ( {|h_{se} |^2  > \Delta } ) + \pi _1 \Pr ( {|h_{se} |^2  - |h_{pe} |^2 \gamma _p \Delta  > \Delta } ),
 \end{split}
\end{equation}
where the second equality is obtained by using $C_{se}$ from (5). Noting that RVs $|{h_{se}}{|^2}$ and $|{h_{pe}}|^2$ are exponentially distributed and independent of each other, we can express the terms $\Pr ( {|h_{se} |^2  > \Delta } )$ and $\Pr ( {|h_{se} |^2  - |h_{pe} |^2 \gamma _p \Delta  > \Delta } )$ as
\begin{equation}\label{equa38}
\Pr ( {|h_{se} |^2  > \Delta } ) = \exp ( - \frac{\Delta }{{\sigma _{se}^2 }}),
\end{equation}
and
\begin{equation}\label{equa39}
\Pr ( {|h_{se} |^2  - |h_{pe} |^2 \gamma _p \Delta  > \Delta } ) = \frac{{\sigma _{se}^2 }}{{\sigma _{pe}^2 \gamma _p \Delta  + \sigma _{se}^2 }}\exp ( - \frac{\Delta }{{\sigma _{se}^2 }}),
\end{equation}
where $\sigma _{se}^2 $ and $\sigma _{pe}^2$ are the expected values of RVs $|h_{se} |^2$ and $|h_{pe} |^2$, respectively.

\subsection{Single-Relay Selection}
In this subsection, we present the SRT analysis of the proposed SRS scheme. Given $\hat H = H_0 $, the OP of the cognitive transmission relying on SRS is given by
\begin{equation}\label{equa40}
\begin{split}
P_{{\textrm{out}}}^{{\textrm{single}}}  =& \Pr ({C_{bd}  < R, {{\cal D} = \emptyset } | {\hat H  = H_0 } } ) \\
&+ \sum\limits_{n = 1}^{2^N  - 1} {\Pr ({C_{bd}  < R, {\cal D} = {\cal D}_n | {\hat H  = H_0 } } )},
 \end{split}
\end{equation}
where $C_{bd}  $ represents the capacity of the channel from the ``best" SR to SD. In the case of ${\cal D} = \emptyset$, no SR is chosen to forward the source signal, which leads to ${C_{bd} } = 0$ for ${\cal D} = \emptyset$. Substituting this result into (40) gives
\begin{equation}\label{equa41}
\begin{split}
P_{{\textrm{out}}}^{{\textrm{single}}}  =& \Pr ( {{\cal D} = \emptyset } {| {\hat H  = H_0 } } ) \\
&+ \sum\limits_{n = 1}^{2^N  - 1} {\Pr ({C_{bd}  < R, {\cal D} = {\cal D}_n | {\hat H  = H_0 } } )}.
\end{split}
\end{equation}
Using (2), (9), (10) and (14), we can rewrite (41) as (42) at the top of the following page,
\begin{figure*}
\begin{equation}\label{equa42}
\begin{split}
 P_{{\textrm{out}}}^{{\textrm{single}}}  =& \pi _0 \prod\limits_{i = 1}^N {\Pr ( {|h_{si} |^2  < \Lambda } )}  + \pi _1 \prod\limits_{i = 1}^N {\Pr ( {|h_{si} |^2  < \Lambda |h_{pi} |^2 \gamma _p  + \Lambda } )}\\
&+ \pi _0 \sum\limits_{n = 1}^{2^N  - 1} {\prod\limits_{i \in {\cal D}_n } {\Pr ( {|h_{si} |^2  > \Lambda } )} \prod\limits_{j \in \bar {\cal D}_n } {\Pr ( {|h_{sj} |^2  < \Lambda } )}\Pr ( {\mathop {\max }\limits_{i \in {\cal {\cal D}}_n } |h_{id} |^2  < \Lambda } )}  \\
&+ \pi _1 \sum\limits_{n = 1}^{2^N  - 1} {\prod\limits_{i \in {\cal D}_n } {\Pr ( {|h_{si} |^2  > \Lambda |h_{pi} |^2 \gamma _p  + \Lambda } )} \prod\limits_{j \in \bar {\cal D}_n } {\Pr ( {|h_{sj} |^2  < \Lambda |h_{pj} |^2 \gamma _p  + \Lambda } )}}\\
&\quad\quad\quad\quad\times\Pr ( {\mathop {\max }\limits_{i \in {\cal D}_n } |h_{id} |^2  < \Lambda |h_{pd} |^2 \gamma _p  + \Lambda } ),  \\
\end{split}
\end{equation}
\end{figure*}
where $\Lambda  = \frac{{2^{2R}  - 1}}{{\gamma _s }}$. Noting that $|h_{si} |^2$ and $|h_{pi} |^2$ are independent exponentially distributed random variables with respective means of $\sigma _{si}^2$ and $\sigma _{pi}^2$, we have
\begin{equation}\label{equa43}
\Pr ( {|h_{si} |^2  < \Lambda } ) = 1 - \exp ( - \frac{\Lambda }{{\sigma _{si}^2 }}),
\end{equation}
and
\begin{equation}\label{equa44}
\Pr ( {|h_{si} |^2  < \Lambda |h_{pi} |^2 \gamma _p  + \Lambda } ) = 1 - \frac{{\sigma _{si}^2 }}{{\sigma _{pi}^2 \gamma _p \Lambda  + \sigma _{si}^2 }}\exp ( - \frac{\Lambda }{{\sigma _{si}^2 }}),
\end{equation}
where the terms $\Pr ( {|h_{si} |^2  > \Lambda } )$, $\Pr ( {|h_{sj} |^2  < \Lambda } )$, and $\Pr ( {|h_{si} |^2  > \Lambda |h_{pi} |^2 \gamma _p  + \Lambda } )$ can be similarly determined in closed-form. Moreover, based on Appendix A, we obtain ${\Pr ( {\mathop {\max }\limits_{i \in {\cal D}_n } |h_{id} |^2  < \Lambda } )}$ and ${\Pr ( {\mathop {\max }\limits_{i \in {\cal D}_n } |h_{id} |^2  < \Lambda |h_{pd} |^2 \gamma _p  + \Lambda } )}$ as
\begin{equation}\label{equa45}
\Pr ( {\mathop {\max }\limits_{i \in {\cal D}_n } |h_{id} |^2  < \Lambda } )  = \prod\limits_{i \in {\cal D}_n } {\left[ {1 - \exp ( - \frac{\Lambda }{{\sigma _{id}^2 }})} \right]},
\end{equation}
and
\begin{equation}\label{equa46}
\begin{split}
&\Pr ( {\mathop {\max }\limits_{i \in {\cal{D}}_n } |h_{id} |^2  < \Lambda |h_{pd} |^2 \gamma _p  + \Lambda } )\\
&=1 + \sum\limits_{m = 1}^{2^{|{\cal{D}}_n |}  - 1} {( - 1)^{|\tilde {\cal{D}}_n (m)|}  \exp ( - \sum\limits_{i \in \tilde {\cal{D}}_n (m)} {\frac{\Lambda }{{\sigma _{id}^2 }}} )}\\
&\quad\quad\quad\quad \quad \times (1 + \sum\limits_{i \in \tilde {\cal{D}}_n (m)} {\frac{{\Lambda \gamma _p \sigma _{pd}^2 }}{{\sigma _{id}^2 }}} )^{ - 1},
\end{split}
\end{equation}
where ${\tilde {\cal D}_n (m)}$ represents the $m$-th non-empty subset of ${\cal{D}}_n$. Additionally, the IP of the SRS scheme can be expressed as
\begin{equation}\label{equa47}
\begin{split}
P_{{\textrm{int}}}^{{\textrm{single}}}  =& \Pr (  {C_{be}  > R, {{\cal D} = \emptyset } | {\hat H  = H_0 } } )\\
&+ \sum\limits_{n = 1}^{2^N  - 1} {\Pr ( {C_{be}  > R, {\cal D} = {\cal D}_n | {\hat H  = H_0 } } )},
 \end{split}
\end{equation}
where $C_{be}  $ represents the capacity of the channel spanning from the ``best" SR to E. Given ${\cal D} = \emptyset$, we have ${C_{be} } = 0$, since no relay is chosen for forwarding the source signal. Thus, substituting this result into (47) and using (2), (9), (10) and (16), we arrive at
\begin{equation}\label{equa48}
\begin{split}
 P_{{\textrm{int}}}^{{\textrm{single}}}  =& \pi _0 \sum\limits_{n = 1}^{2^N  - 1} \prod\limits_{i \in {\cal D}_n } {\Pr ( {|h_{si} |^2  > \Lambda } )} \prod\limits_{j \in \bar {\cal D}_n } {\Pr ( {|h_{sj} |^2  < \Lambda } )}\\
 &\quad\quad\quad\quad\times \Pr ( { |h_{be} |^2  > \Lambda } )  \\
&+ \pi _1 \sum\limits_{n = 1}^{2^N  - 1} \prod\limits_{i \in {\cal D}_n } {\Pr ( {|h_{si} |^2  > \Lambda |h_{pi} |^2 \gamma _p  + \Lambda } )} \\ &\quad\quad\quad\quad\times\prod\limits_{j \in \bar {\cal D}_n } {\Pr ( {|h_{sj} |^2  < \Lambda |h_{pj} |^2 \gamma _p  + \Lambda } )}\\
&\quad\quad\quad\quad\times\Pr ( { |h_{be} |^2  > \Lambda |h_{pe} |^2 \gamma _p  + \Lambda } ), \\
\end{split}
\end{equation}
where the closed-form expressions of $\Pr ( {|h_{si} |^2  > \Lambda } )$ and $\Pr ( {|h_{si} |^2  > \Lambda |h_{pi} |^2 \gamma _p  + \Lambda } )$ can be readily obtained by using (43) and (44). Using the results in Appendix B, we can express $\Pr ( { |h_{be} |^2  > \Lambda } )$ and $\Pr (|h_{be} |^2  > \Lambda |h_{pe} |^2 \gamma _p  + \Lambda )$ as
\begin{equation}\label{equa49}
\begin{split}
&\Pr (|h_{be} |^2  > \Lambda ) =  \sum\limits_{i \in {\cal D}_n } \exp ( - \frac{\Lambda }{{\sigma _{ie}^2 }})\\
&\quad\quad\times\left[ {1 + \sum\limits_{m = 1}^{2^{|{\cal D}_n | - 1}  - 1} {( - 1)^{|{\cal C}_n (m)|} (1 + \sum\limits_{j \in {\cal C}_n (m)} {\frac{{\sigma _{id}^2 }}{{\sigma _{jd}^2 }}} )^{ - 1} } } \right],
\end{split}
\end{equation}
and
\begin{equation}\label{equa50}
\begin{split}
&\Pr (|h_{be} |^2  > \Lambda |h_{pe} |^2 \gamma _p  + \Lambda ) = \sum\limits_{i \in {\cal D}_n } {\frac{{\sigma _{ie}^2 }}{{\sigma _{pe}^2 \gamma _p \Lambda  + \sigma _{ie}^2 }}\exp ( - \frac{\Lambda }{{\sigma _{ie}^2 }})}\\
&\quad \quad \times \left[ {1 + \sum\limits_{m = 1}^{2^{|{\cal D}_n | - 1}  - 1} {( - 1)^{|{\cal C}_n (m)|} (1 + \sum\limits_{j \in {\cal C}_n (m)} {\frac{{\sigma _{id}^2 }}{{\sigma _{jd}^2 }}} )^{ - 1} } } \right],
\end{split}
\end{equation}
where ${{\cal C}_n (m)}$ represents the $m$-th non-empty subset of ${{\cal{D}}_n}-\{i\}$ and `$-$' represents the set difference.

\subsection{Multi-Relay Selection}
This subsection analyzes the SRT of our MRS scheme for transmission over Rayleigh fading channels. Similarly to (41), the OP in this case is given by
\begin{equation}\label{equa51}
\begin{split}
P_{{\textrm{out}}}^{{\textrm{multi}}}  &= \Pr ( {\left. {{\cal{D}} = \emptyset } \right|\hat H = H_0 } ) \\
&+ \sum\limits_{n = 1}^{2^N  - 1} {\Pr ( { {C_d^{{\textrm{multi}}}  < R,{\cal{D}} = {\cal{D}}_n } |\hat H = H_0 } )}.
\end{split}
\end{equation}
Using (2), (9), (10) and (23), we can rewrite (51) as (52) at the top of the following page,
\begin{figure*}
\begin{equation}\label{equa52}
\begin{split}
 P_{{\textrm{out}}}^{{\textrm{multi}}}  =& \pi _0 \prod\limits_{i = 1}^N {\Pr ( {|h_{si} |^2  < \Lambda } )}  + \pi _1 \prod\limits_{i = 1}^N {\Pr ( {|h_{si} |^2  < \Lambda |h_{pi} |^2 \gamma _p  + \Lambda } )}  \\
&+ \pi _0 \sum\limits_{n = 1}^{2^N  - 1} {\prod\limits_{i \in {\cal{D}}_n } {\Pr ( {|h_{si} |^2  > \Lambda } )} \prod\limits_{j \in \bar {\cal{D}}_n } {\Pr ( {|h_{sj} |^2  < \Lambda } )} \Pr ( {\sum\limits_{i \in {\cal{D}}_n } {|h_{id} |^2 }  < \Lambda } )}  \\
&+ \pi _1 \sum\limits_{n = 1}^{2^N  - 1} {\prod\limits_{i \in {\cal{D}}_n } {\Pr ( {|h_{si} |^2  > \Lambda |h_{pi} |^2 \gamma _p  + \Lambda } )} \prod\limits_{j \in \bar {\cal{D}}_n } {\Pr ( {|h_{sj} |^2  < \Lambda |h_{pj} |^2 \gamma _p  + \Lambda } )} }  \\
&\quad\quad\quad\quad\quad \times \Pr ( {\sum\limits_{i \in {\cal{D}}_n } {|h_{id} |^2 }  < \gamma _p \Lambda |h_{pd} |^2  + \Lambda } ), \\
\end{split}
\end{equation}
\end{figure*}
where the closed-form expressions of $\Pr ( {|h_{si} |^2  < \Lambda } )$, $\Pr ( {|h_{si} |^2  < \Lambda |h_{pi} |^2 \gamma _p  + \Lambda } )$, $\Pr ( {|h_{si} |^2  > \Lambda } )$, $\Pr ( {|h_{sj} |^2  < \Lambda } )$ and $\Pr ( {|h_{si} |^2  > \Lambda |h_{pi} |^2 \gamma _p  + \Lambda } )$ can be readily derived, as shown in (43) and (44). However, it is challenging to obtain the closed-form expressions of $\Pr (\sum\limits_{i \in {\cal{D}}_n } {|h_{id} |^2 }  < \Lambda )$ and $\Pr ( {\sum\limits_{i \in {\cal{D}}_n } {|h_{id} |^2 }  < \gamma _p \Lambda |h_{pd} |^2  + \Lambda } )$. For simplicity, we assume that the fading coefficients of all SRs-SD channels, i.e. $|h_{id}|^2$ for $i \in \{1,2,\cdots,N\}$, are i.i.d. RVs having the same mean (average channel gain) denoted by $\sigma^2_{d}=E(|h_{id}|^2)$. This assumption is widely used in the cooperative relaying literature and it is valid in a statistical sense, provided that all SRs are uniformly distributed over a certain geographical area. Assuming that RVs of $|h_{id}|^2$ for $i \in {\cal{D}}_n$ are i.i.d., based on Appendix C, we arrive at
\begin{equation}\label{equa53}
\Pr (\sum\limits_{i \in {\cal{D}}_n } {|h_{id} |^2 }  < \Lambda ) = \Gamma ( {\frac{{\Lambda}}{{ \sigma _d^2 }},|{\cal{D}}_n |} ),
\end{equation}
and
\begin{equation}\label{equa54}
\begin{split}
& \Pr (\sum\limits_{i \in {\cal D}_n } {|h_{id} |^2 }  < \gamma _p \Lambda |h_{pd} |^2  + \Lambda ) =\Gamma (\frac{{\Lambda }}{{\sigma _d^2 }},|{\cal D}_n |)\\
&\quad \quad + \frac{{[1 - \Gamma (\Lambda \sigma _d^{ - 2}  + \sigma _{pd}^{ - 2} \gamma _p^{ - 1} ,|{\cal{D}}_n |)]}}{{(1 + \sigma _d^2 \sigma _{pd}^{ - 2} \gamma _p^{ - 1} \Lambda ^{ - 1} )^{|{\cal{D}}_n |} }}e^{1/(\sigma _{pd}^2 \gamma _p )}, \\
 \end{split}
\end{equation}
where $\Gamma (x,k) = \int_0^x {\frac{{t^{k - 1} }}{{\Gamma (k)}}e^{ - t} dt}$ is known as the incomplete Gamma function [32]. Substituting (53) and (54) into (52) yields a closed-form OP expression for the proposed MRS scheme.

Next, we present the IP analysis of the MRS scheme. Similarly to (48), the IP of the MRS can be obtained from (24) as
\begin{equation}\label{equa55}
\begin{split}
 P_{{\textrm{int}}}^{{\textrm{multi}}}  =& \pi _0 \sum\limits_{n = 1}^{2^N  - 1} \prod\limits_{i \in {\cal D}_n } {\Pr ( {|h_{si} |^2  > \Lambda } )} \prod\limits_{j \in \bar {\cal D}_n } {\Pr ( {|h_{sj} |^2  < \Lambda } )} \\
 &\quad\quad\quad\times \Pr (\frac{{|{{H}}_d^H  {{H}}_e |^2 }}{{|{{H}}_d |^2 }} > \Lambda )  \\
&+ \pi _1 \sum\limits_{n = 1}^{2^N  - 1} \prod\limits_{i \in {\cal D}_n } {\Pr ( {|h_{si} |^2  > \Lambda |h_{pi} |^2 \gamma _p  + \Lambda } )}\\ &\quad\quad\quad\quad\times \prod\limits_{j \in \bar {\cal D}_n } {\Pr ( {|h_{sj} |^2  < \Lambda |h_{pj} |^2 \gamma _p  + \Lambda } )}   \\
&\quad\quad\quad\quad\times \Pr (\frac{{|{{H}}_d^H  {{H}}_e |^2 }}{{|{{H}}_d |^2 }} > \gamma _p \Lambda |h_{pe} |^2  + \Lambda ), \\
 \end{split}
\end{equation}
where the closed-form expressions of $\Pr ( {|h_{si} |^2  > \Lambda } )$, $\Pr ( {|h_{sj} |^2  < \Lambda } )$, $\Pr ( {|h_{si} |^2  > \Lambda |h_{pi} |^2 \gamma _p  + \Lambda } )$ and $\Pr ( {|h_{sj} |^2  < \Lambda |h_{pj} |^2 \gamma _p  + \Lambda } )$ may be readily derived by using (43) and (44). However, it is challenging to obtain the closed-form solutions for $\Pr (\frac{{|{{H}}_d^H  {{H}}_e |^2 }}{{|{{H}}_d |^2 }} > \Lambda )$ and $\Pr (\frac{{|{{H}}_d^H  {{H}}_e |^2 }}{{|{{H}}_d |^2 }} > \gamma _p \Lambda |h_{pe} |^2  + \Lambda )$. Although finding a general closed-form IP expression for the MRS scheme is challenging, we can obtain the numerical IP results with the aid of computer simulations.

\section{Numerical Results and Discussions}
In this section, we present our performance comparisons among the direct transmission, the SRS and MRS schemes in terms of their SRT. To be specific, the analytic IP versus OP of the three schemes are obtained by plotting (33), (37), (42), (48), (52) and (55). {The simulated IP and OP results of the three schemes are also given to verify the correctness of the theoretical SRT analysis. In our computer simulations, the fading amplitudes (e.g., $|h_{sd}|$, $|h_{si}|$, $|h_{id}|$, etc.) are first generated based on the Rayleigh distribution having different variances for different channels. Then, the randomly generated fading amplitudes are substituted into the definition of an outage (or intercept) event, which would determine whether an outage (or intercept) event occurs or not. By repeatedly achieving this process, we can calculate the relative frequency of occurrence for an outage (intercept) event, which is the simulated OP (or IP). Additionally, the SDP $P_d$ and FAP $P_f$ are set to $P_d=0.99$ and $P_f=0.01$, unless otherwise stated. The primary signal-to-noise ratio (SNR) of $\gamma_p=10{\textrm{dB}}$ and the data rate of $R=1{\textrm{bit/s/Hz}}$ are used in our numerical evaluations.}
\begin{figure}
  \centering
  {\includegraphics[scale=0.6]{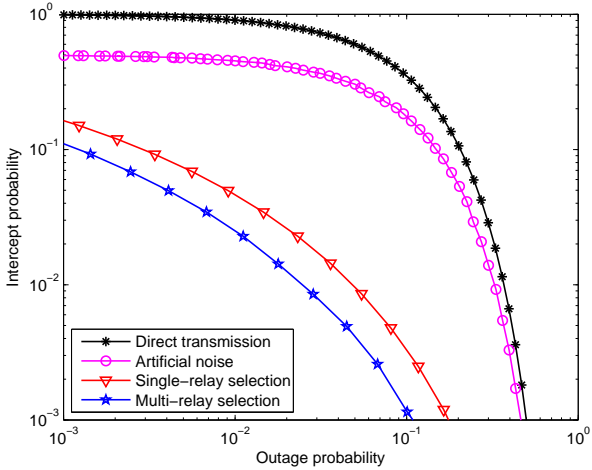}\\
  \caption{{IP versus OP of the direct transmission, the SRS and the MRS schemes for different $P_0$ with $P_0=0.8$, $\gamma _s \in [0,35{\textrm{dB}}]$, $N=6$, $\sigma^2_{sd}=\sigma^2_{si}=\sigma^2_{id}=1$, $\sigma^2_{se}=\sigma^2_{ie}=0.1$, and $\sigma^2_{pd}=\sigma^2_{pe}=\sigma^2_{pi}=0.2$.}}\label{Fig3}}
\end{figure}

{The artificial noise based method [35], [36] is also considered for the purpose of numerical comparison with the relay selection schemes. To be specific, in the artificial noise based scheme, ST directly transmits its signal $x_s$ to SD, while $N$ SRs attempt to confuse the eavesdropper by sending an interfering signal (referred to as artificial noise) that is approximately designed to lie in the null-space of the legitimate main channel. In this way, the artificial noise will impose interference on the eavesdropper without affecting the SD. For a fair comparison, the total transmit power of the desired signal $x_s$ and the artificial noise are constrained to $P_s$. Moreover, the equal power allocation method [35] is used in the numerical evaluation.}

Fig. 3 shows the IP versus OP of the direct transmission, as well as the SRS and MRS schemes for $P_0=0.8$, where the solid lines and discrete marker symbols represent the analytic and simulated results, respectively. It can be seen from Fig. 3 that the IP of the direct transmission, {the artificial noise based} as well as of the proposed SRS and MRS schemes all improve upon tolerating a higher OP, implying that a trade-off exists between the IP (security) and the OP (reliability) of CR transmissions. Fig. 3 also shows that both the proposed SRS and MRS schemes outperform the direct transmission {and the artificial noise based approaches} in terms of their SRT{, showing the advantage of exploiting relay selection against the eavesdropping attack}. Moreover, the SRT performance of the MRS is better than that of the SRS. Although the MRS achieves a better SRT performance than its SRS-aided counterpart, this result is obtained at the cost of a higher implementation complexity, since multiple SRs require high-complexity symbol-level synchronization for simultaneously transmitting to the SD, whereas the SRS does not require such elaborate synchronization.

\begin{figure}
  \centering
  {\includegraphics[scale=0.6]{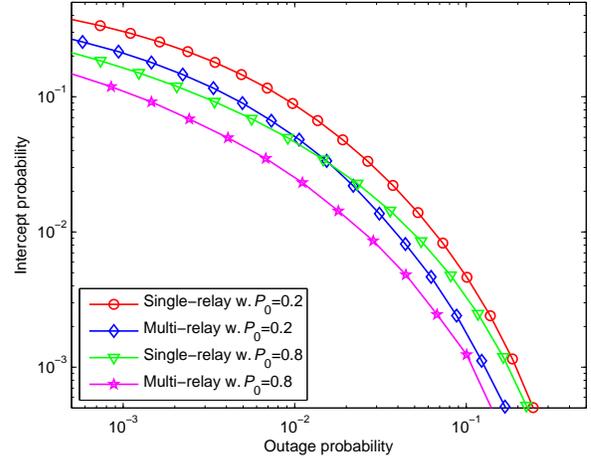}\\
  \caption{IP versus OP of the SRS and MRS schemes for different $P_0$ with $\gamma _s  \in [0,30{\textrm{dB}}]$, $N=6$, $\sigma^2_{sd}=\sigma^2_{si}=\sigma^2_{id}=1$, $\sigma^2_{se}=\sigma^2_{ie}=0.1$, and $\sigma^2_{pd}=\sigma^2_{pe}=\sigma^2_{pi}=0.2$.}\label{Fig4}}
\end{figure}
Fig. 4 illustrates our numerical SRT comparison between the SRS and MRS schemes for {$P_0=0.2$ and $P_0=0.8$}. Observe from Fig. 4 that the MRS scheme performs better than the SRS in terms of its SRT performance for both $P_0=0.2$ and $P_0=0.8$. It is also seen from Fig. 4 that as $P_0$ increases from $0.2$ to $0.8$, the SRT of both the SRS and MRS schemes improves. This is because upon increasing $P_0$, the licensed band becomes unoccupied by the PUs with a higher probability and hence the secondary users (SUs) have more opportunities for accessing the licensed band for their data transmissions, which leads to a reduction of the OP for CR transmissions. {{Meanwhile, increasing $P_0$ may simultaneously result in an increase of the IP, since the eavesdropper also has more opportunities for tapping the cognitive transmissions. However, in both the SRS and MRS schemes, the relay selection is performed for the sake of maximizing the legitimate transmission capacity without affecting the eavesdropper's channel capacity. Hence, upon increasing $P_0$, it becomes more likely that the reduction of OP is more significant than the increase of IP, hence leading to an overall SRT improvement for the SRS and MRS schemes.}}

\begin{figure}
  \centering
  {\includegraphics[scale=0.6]{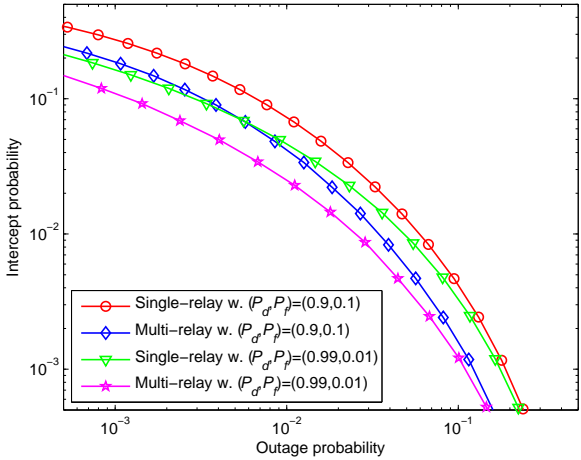}\\
  \caption{IP versus OP of the SRS and the MRS schemes for different $(P_d,P_f)$ with $P_0=0.8$, $\gamma _s  \in [0,30{\textrm{dB}}]$, $N=6$, $\sigma^2_{sd}=\sigma^2_{si}=\sigma^2_{id}=1$, $\sigma^2_{se}=\sigma^2_{ie}=0.1$, and $\sigma^2_{pd}=\sigma^2_{pe}=\sigma^2_{pi}=0.2$.}\label{Fig5}}
\end{figure}
In Fig. 5, we depict the IP versus OP of the SRS and MRS schemes for different spectrum sensing reliabilities, where $(P_d,P_f)=(0.9,0.1)$ and $(P_d,P_f)=(0.99,0.01)$ are considered. It is observed that as the spectrum sensing reliability is improved from $(P_d,P_f)=(0.9,0.1)$ to $(P_d,P_f)=(0.99,0.01)$, the SRTs of the SRS and MRS schemes improve accordingly. {This is due to the fact that for an improved sensing reliability, an unoccupied licensed band would be detected more accurately and hence less mutual interference occurs between the PUs and SUs, which results in a better SRT for the secondary transmissions.} Fig. 5 also shows that for $(P_d,P_f)=(0.9,0.1)$ and $(P_d,P_f)=(0.99,0.01)$, the MRS approach outperforms the SRS scheme in terms of the SRT, which further confirms the advantage of the MRS for protecting the secondary transmissions against eavesdropping attacks.

\begin{figure}
  \centering
  {\includegraphics[scale=0.6]{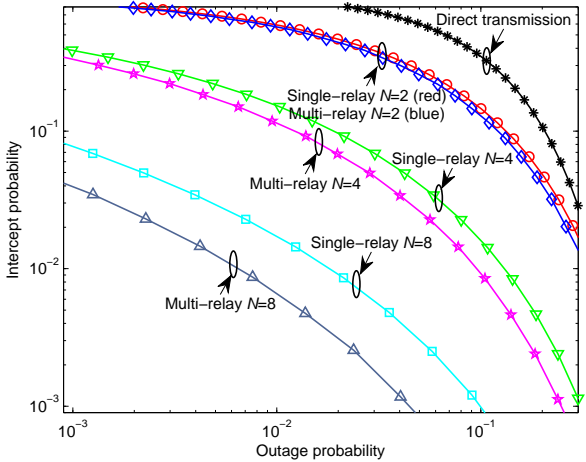}\\
  \caption{{IP versus OP of the direct transmission, the SRS and the MRS schemes for different $N$ with $P_0=0.8$, $\gamma _s  \in [0,30{\textrm{dB}}]$, $\sigma^2_{sd}=\sigma^2_{si}=\sigma^2_{id}=1$, $\sigma^2_{se}=\sigma^2_{ie}=0.1$, and $\sigma^2_{pd}=\sigma^2_{pe}=\sigma^2_{pi}=0.2$.}}\label{Fig6}}
\end{figure}
Fig. 6 shows the IP versus OP of the conventional direct transmission as well as of the proposed SRS and MRS schemes for {$N=2$}, $N=4$, and $N=8$. It is seen from Fig. 6 that the SRTs of the proposed SRS and MRS schemes are generally better than that of the conventional direct transmission for $N=2$, $N=4$ and $N=8$. Moreover, as the number of SRs increases from $N=2$ to $8$, the SRT of the SRS and MRS schemes significantly improves, explicitly demonstrating the security and reliability benefits of exploiting multiple SRs for assisting the secondary transmissions. In other words, the security and reliability of the secondary transmissions can be concurrently improved by increasing the number of SRs. Additionally, {as shown in Fig. 6, upon increasing the number of SRs from $N=2$ to $8$, the SRT improvement of MRS over SRS becomes more notable}. Again, the SRT advantage of the MRS over the SRS comes at the expense of requiring elaborate symbol-level synchronization among the multiple SRs for simultaneously transmitting to the SD.

\section{Conclusions}
In this paper, we proposed relay selection schemes for a CR network consisting of a ST, a SD and multiple SRs communicating in the presence of an eavesdropper. We examined the SRT performance of the SRS and MRS assisted secondary transmissions in the presence of realistic spectrum sensing, where both the security and reliability of secondary transmissions are characterized in terms of their IP and OP, respectively. We also analyzed the SRT of the conventional direct transmission as a benchmark. It was illustrated that as the spectrum sensing reliability increases, the SRTs of both the SRS and MRS schemes improve. We also showed that the proposed SRS and MRS schemes generally outperform the conventional direct transmission and artificial noise based approaches in terms of their SRT. Moreover, the SRT performance of MRS is better than that of SRS. Additionally, as the number of SRs increases, the SRTs of both the SRS and of the MRS schemes improve significantly, demonstrating their benefits in terms of enhancing both the security and reliability of secondary transmissions.

\appendices
\section{Derivation of (45) and (46)}
Letting $|h_{id} |^2  = x_i $ and $|h_{pd} |^2  = y$, the left hand side of (45) and (46) can be rewritten as $\Pr ( {\mathop {\max }\limits_{i \in {\cal {D}}_n } x_i  < \Lambda } )$ and $\Pr ( {\mathop {\max }\limits_{i \in {\cal {D}}_n } x_i  < \Lambda \gamma _p y + \Lambda } )$, respectively. Noting that random variables $|h_{id} |^2$ and $|h_{pd} |^2$ are exponentially distributed with respective means $\sigma _{id}^2$ and $\sigma _{pd}^2$, and independent of each other, we obtain
\begin{equation}\nonumber
\begin{split}
\Pr (\mathop {\max }\limits_{i \in {\cal D}_n } x_i  < \Lambda ) &= \prod\limits_{i \in {\cal D}_n } {\Pr ( {|h_{id} |^2  < \Lambda } )} \\
&= \prod\limits_{i \in {\cal D}_n } {\left[ {1 - \exp ( - \frac{\Lambda }{{\sigma _{id}^2 }})} \right]},\\
\end{split}\label{A.1}\tag{A.1}
\end{equation}
which is (45). Similarly, the term $\Pr ( {\mathop {\max }\limits_{i \in {\cal {D}}_n } x_i  < \Lambda \gamma _p y + \Lambda } )$ can be computed as
\begin{equation}\nonumber
\begin{split}
&\Pr ( {\mathop {\max }\limits_{i \in {\cal D}_n } x_i  < \Lambda \gamma _p y + \Lambda } ) \\
&= \int_0^\infty  {\frac{1}{{\sigma _{pd}^2 }}\exp ( - \frac{y}{{\sigma _{pd}^2 }})\prod\limits_{i \in {\cal D}_n } {( {1 - \exp ( - \frac{{\Lambda \gamma _p y + \Lambda }}{{\sigma _{id}^2 }})} )} dy},
\end{split}\label{A.2}\tag{A.2}
\end{equation}
wherein $\prod\limits_{i \in {\cal D}_n } {( {1 - \exp ( - \frac{{\Lambda \gamma _p y + \Lambda }}{{\sigma _{id}^2 }})} )}$ can be further expanded as
\begin{equation}\nonumber
\begin{split}
&\prod\limits_{i \in {\cal D}_n } {( {1 - \exp ( - \frac{{\Lambda \gamma _p y + \Lambda }}{{\sigma _{id}^2 }})} )}  = 1 \\
&+ \sum\limits_{m = 1}^{2^{|{\cal D}_n |}  - 1} {( - 1)^{|\tilde {\cal D}_n (m)|} \exp ( - \sum\limits_{i \in \tilde {\cal D}_n (m)} {\frac{{\Lambda \gamma _p y + \Lambda }}{{\sigma _{id}^2 }}} )},
\end{split}\label{A.3}\tag{A.3}
\end{equation}
where ${|{\cal{D}}_n |}$ is the cardinality of set ${\cal{D}}_n$, ${\tilde {\cal D}_n (m)}$ represents the $m$-th non-empty subset of ${\cal{D}}_n$, and $|{\tilde {\cal D}_n (m)}|$ is the cardinality of set ${\tilde {\cal D}_n (m)}$. Substituting $\prod\limits_{i \in {\cal D}_n } {( {1 - \exp ( - \frac{{\Lambda \gamma _p y + \Lambda }}{{\sigma _{id}^2 }})} )}$ from (A.3) into (A.2) yields
\begin{equation}\nonumber
\begin{split}
& \Pr ( {\mathop {\max }\limits_{i \in {\cal D}_n } x_i  < \Lambda \gamma _p y + \Lambda } ) = \int_0^\infty  {\frac{1}{{\sigma _{pd}^2 }}\exp ( - \frac{y}{{\sigma _{pd}^2 }})dy}  \\
& + \sum\limits_{m = 1}^{2^{|{\cal D}_n |}  - 1} ( - 1)^{|\tilde {\cal D}_n (m)|} \frac{1}{{\sigma _{pd}^2 }}\\
&\quad\quad\times\int_0^\infty  {\exp ( - \frac{y}{{\sigma _{pd}^2 }} - \sum\limits_{i \in \tilde {\cal D}_n (m)} {\frac{{\Lambda \gamma _p y + \Lambda }}{{\sigma _{id}^2 }}} )dy} .  \\
\end{split}\label{A.4}\tag{A.4}
\end{equation}
Finally, performing the integration of (A.4) yields
\begin{equation}\nonumber
\begin{split}
&\Pr ( {\mathop {\max }\limits_{i \in {\cal {\cal D}}_n } x_i  < \Lambda \gamma _p y + \Lambda } ) = 1 \\
&\quad + \sum\limits_{m = 1}^{2^{|{\cal D}_n |}  - 1} {( - 1)^{|\tilde {\cal D}_n (m)|}  \exp ( - \sum\limits_{i \in \tilde {\cal D}_n (m)} {\frac{\Lambda }{{\sigma _{id}^2 }}} )} \\
&\quad\times(1 + \sum\limits_{i \in \tilde {\cal D}_n (m)}{\frac{{\Lambda \gamma _p \sigma _{pd}^2 }}{{\sigma _{id}^2 }}} )^{ - 1}. \\
\end{split}\label{A.5}\tag{A.5}
\end{equation}
This completes the proof of (45) and (46).

\section{Proof of (49) and (50)}
Given ${\cal D} = {\cal D}_n$, any SR within ${\cal D}_n$ can be selected as the ``best" relay for forwarding the source signal. Thus, using the law of total probability, we have
\begin{equation}\nonumber
\begin{split}
&\Pr (|h_{be} |^2  > \Lambda ) = \sum\limits_{i \in {\cal D}_n } {\Pr (|h_{ie} |^2  > \Lambda ,b = i)}  \\
&= \sum\limits_{i \in {\cal D}_n } {\Pr (|h_{ie} |^2  > \Lambda ,|h_{id} |^2  > \mathop {\max }\limits_{j \in { {\cal D}_n - \{i\}} } |h_{jd} |^2 )}  \\
&= \sum\limits_{i \in {\cal D}_n } {\Pr (|h_{ie} |^2  > \Lambda )\Pr (\mathop {\max }\limits_{j \in { {\cal D}_n - \{i\}} } |h_{jd} |^2  < |h_{id} |^2 )},  \\
\end{split}\label{B.1}\tag{B.1}
\end{equation}
where in the first line, variable `$b$' stands for the best SR and the second equality is obtained from (13) and `$-$' represents the set difference. Noting that ${|h_{ie} |^2 }$ is an exponentially distributed random variable with a mean of ${\sigma _{ie}^2 }$, we obtain
\begin{equation}\nonumber
\Pr (|h_{ie} |^2  > \Lambda ) = \exp ( - \frac{\Lambda }{{\sigma _{ie}^2 }}).\label{B.2}\tag{B.2}
\end{equation}
Letting $|h_{jd} |^2  = x_j$ and $|h_{id} |^2  = y$, we have
\begin{equation}\nonumber
\begin{split}
&\Pr (\mathop {\max }\limits_{j \in { {\cal D}_n - \{i\}} } |h_{jd} |^2  < |h_{id} |^2 ) \\
&= \int_0^\infty  {\frac{1}{{\sigma _{id}^2 }}\exp ( - \frac{y}{{\sigma _{id}^2 }})\prod\limits_{j \in { {\cal D}_n - \{i\}} } {( {1 - \exp ( - \frac{y}{{\sigma _{jd}^2 }})} )} dy},
\end{split}\label{B.3}\tag{B.3}
\end{equation}
wherein $\prod\limits_{j \in { {\cal D}_n - \{i\}} } {( {1 - \exp ( - \frac{y}{{\sigma _{jd}^2 }})} )}$ is expanded by
\begin{equation}\nonumber
\begin{split}
&\prod\limits_{j \in { {\cal D}_n - \{i\}} } {( {1 - \exp ( - \frac{y}{{\sigma _{jd}^2 }})} )}  = 1 \\
&\quad+ \sum\limits_{m = 1}^{2^{|{\cal D}_n | - 1}  - 1} {( - 1)^{|{\cal C}_n (m)|} \exp ( - \sum\limits_{j \in {\cal C}_n (m)} {\frac{y}{{\sigma _{jd}^2 }}} )},
\end{split}\label{B.4}\tag{B.4}
\end{equation}
where ${|{\cal{D}}_n |}$ denotes the cardinality of the set ${\cal{D}}_n$ and ${{\cal C}_n (m)}$ represents the $m$-th non-empty subset of ``${{\cal{D}}_n}-\{i\}$". Combining (B.3) and (B.4), we obtain
\begin{equation}\nonumber
\begin{split}
&\Pr (\mathop {\max }\limits_{j \in { {\cal D}_n - \{i\}} } |h_{jd} |^2  < |h_{id} |^2 ) = 1 \\
&\quad + \sum\limits_{m = 1}^{2^{|{\cal D}_n | - 1}  - 1} {( - 1)^{|{\cal C}_n (m)|} (1 + \sum\limits_{j \in {\cal C}_n (m)} {\frac{{\sigma _{id}^2 }}{{\sigma _{jd}^2 }}} )^{ - 1} }.
\end{split}\label{B.5}\tag{B.5}
\end{equation}
Substituting (B.2) and (B.5) into (B.1) gives (B.6) at the top of the following page,
\begin{figure*}
\begin{equation}\nonumber
\Pr (|h_{be} |^2  > \Lambda ) = \sum\limits_{i \in {\cal D}_n } {\exp ( - \frac{\Lambda }{{\sigma _{ie}^2 }})\left[ {1 + \sum\limits_{m = 1}^{2^{|{\cal D}_n | - 1}  - 1} {( - 1)^{|{\cal C}_n (m)|} (1 + \sum\limits_{j \in {\cal C}_n (m)} {\frac{{\sigma _{id}^2 }}{{\sigma _{jd}^2 }}} )^{ - 1} } } \right]},
\label{B.6}\tag{B.6}
\end{equation}
\end{figure*}
which is (49). Similarly to (B.1), we can rewrite $\Pr (|h_{be} |^2  > \Lambda |h_{pe} |^2 \gamma _p  + \Lambda )$ as
\begin{equation}\nonumber
\begin{split}
&\Pr (|h_{be} |^2  > \Lambda |h_{pe} |^2 \gamma _p  + \Lambda ) \\
&=  \sum\limits_{i \in {\cal D}_n } {\Pr (|h_{ie} |^2  > \Lambda |h_{pe} |^2 \gamma _p  + \Lambda )}\\
&\quad\quad\times\Pr (\mathop {\max }\limits_{j \in \{ {\cal D}_n  - i\} } |h_{jd} |^2  < |h_{id} |^2 ) . \\
\end{split}\label{B.7}\tag{B.7}
\end{equation}
Since the random variables $|h_{ie} |^2$ and $|h_{pe} |^2$ are independently and exponentially distributed with respective means of ${\sigma _{ie}^2 }$ and ${\sigma _{pe}^2 }$, we readily arrive at
\begin{equation}\nonumber
\Pr (|h_{ie} |^2  > \Lambda |h_{pe} |^2 \gamma _p  + \Lambda ) = \frac{{\sigma _{ie}^2 }}{{\sigma _{pe}^2 \gamma _p \Lambda  + \sigma _{ie}^2 }}\exp ( - \frac{\Lambda }{{\sigma _{ie}^2 }}).
\label{B.8}\tag{B.8}
\end{equation}
Substituting (B.5) and (B.8) into (B.7) gives (B.9) at the top of the following page,
\begin{figure*}
\begin{equation}\nonumber
\begin{split}
\Pr (|h_{be} |^2  > \Lambda |h_{pe} |^2 \gamma _p  + \Lambda ) = \sum\limits_{i \in {\cal D}_n } {\frac{{\sigma _{ie}^2 }}{{\sigma _{pe}^2 \gamma _p \Lambda  + \sigma _{ie}^2 }}\exp ( - \frac{\Lambda }{{\sigma _{ie}^2 }})}\left[ {1 + \sum\limits_{m = 1}^{2^{|{\cal D}_n | - 1}  - 1} {( - 1)^{|{\cal C}_n (m)|} (1 + \sum\limits_{j \in {\cal C}_n (m)} {\frac{{\sigma _{id}^2 }}{{\sigma _{jd}^2 }}} )^{ - 1} } } \right],
\end{split}
\label{B.9}\tag{B.9}
\end{equation}
\end{figure*}
which is (50).

\section{Proof of (53) and (54)}
Upon introducing the notation of $X = \sum\limits_{i \in {\cal{D}}_n } {|h_{id} |^2 }$ and $ Y = |h_{pd} |^2$, we can rewrite the terms $\Pr (\sum\limits_{i \in {\cal{D}}_n } {|h_{id} |^2 }  < \Lambda )$ and $\Pr ( {\sum\limits_{i \in {\cal{D}}_n } {|h_{id} |^2 }  < \gamma _p \Lambda |h_{pd} |^2  + \Lambda } )$ as $\Pr (X < \Lambda )$ and $\Pr (X < \gamma _p \Lambda Y + \Lambda )$, respectively. Noting that the fading coefficients of all SR-SD channels, i.e. $|h_{id}|^2$ for $i \in \{1,2,\cdots,N\}$, are assumed to be i.i.d., we obtain the probability density function (PDF) of $X = \sum\limits_{i \in {\cal{D}}_n } {|h_{id} |^2 }$ as
\begin{equation}
f_X (x) = \frac{1}{{\Gamma (|{\cal{D}}_n |)\sigma _d^{2|{\cal{D}}_n |} }}x^{|{\cal{D}}_n | - 1} \exp ( - \frac{x}{{\sigma _d^2 }}),\label{C.1}\tag{C.1}
\end{equation}
where $\sigma^2_{d}=E(|h_{id}|^2)$. Meanwhile, the random variable $ Y = |h_{pd} |^2$ is exponentially distributed and its PDF is given by
\begin{equation}
f_Y (y) = \frac{1}{{\sigma _{pd}^2 }}\exp ( - \frac{y}{{\sigma _{pd}^2 }}),\label{C.2}\tag{C.2}
\end{equation}
where $\sigma^2_{pd}=E(|h_{pd}|^2)$. Using (C.1), we arrive at
\begin{equation}
\begin{split}
 \Pr (X < \Lambda ) &= \int_0^{\Lambda } {\frac{1}{{\Gamma (|{\cal{D}}_n |)\sigma _d^{2|{\cal{D}}_n |} }}x^{|{\cal{D}}_n | - 1} \exp ( - \frac{x}{{\sigma _d^2 }})dx}  \\
&= \int_0^{\frac{{\Lambda }}{{\sigma _d^2 }}} {\frac{{t^{|{\cal{D}}_n | - 1} }}{{\Gamma (|{\cal{D}}_n |)}}\exp ( - t)dt}  \\
&= \Gamma (\frac{{\Lambda }}{{\sigma _d^2 }},|{\cal{D}}_n |), \\
\end{split}\label{C.3}\tag{C.3}
\end{equation}
where the second equality is obtained by substituting $\frac{x}{{\sigma _d^2 }} = t$ and $\Gamma (a,k) = \int_0^a {\frac{{t^{k - 1} }}{{\Gamma (k)}}\exp ( - t)dt}$ is known as the incomplete Gamma function. Additionally, considering that the random variables $X$ and $Y$ are independent of each other, we obtain $\Pr (X < \gamma _p \Lambda Y + \Lambda )$ as
\begin{equation}
\begin{split}
&\Pr (X < \gamma _p \Lambda Y + \Lambda ) = \int_0^{\Lambda } {f_X (x)dx}  \\
&\quad \quad \quad + \int_{\Lambda }^\infty  {\int_{\frac{x}{{\gamma _p \Lambda }} - \frac{1}{{\gamma _p }}}^\infty  {f_X (x)f_Y (y)dxdy}}.
\end{split}\label{C.4}\tag{C.4}
\end{equation}
Substituting $f_X (x)$ and $f_Y (y)$ from (C.1) and (C.2) into (C.4) yields
\begin{equation}
\begin{split}
&\Pr (X < \gamma _p \Lambda Y + \Lambda ) \\
&= \Gamma (\frac{{\Lambda }}{{\sigma _d^2 }},|{\cal{D}}_n |) + \int_{\Lambda }^\infty  {\frac{{e^{1/(\sigma _{pd}^2 \gamma _p )} x^{|{\cal{D}}_n | - 1} }}{{\Gamma (|{\cal{D}}_n |)\sigma _d^{2|{\cal{D}}_n |} }}\exp ( - \frac{x}{{\sigma _d^2 }} - \frac{x}{{\sigma _{pd}^2 \gamma _p \Lambda }})dx}  \\
&= \Gamma (\frac{{\Lambda }}{{\sigma _d^2 }},|{\cal{D}}_n |) + \frac{{[1 - \Gamma (\Lambda \sigma _d^{ - 2}  + \sigma _{pd}^{ - 2} \gamma _p^{ - 1} ,|{\cal{D}}_n |)]}}{{(1 + \sigma _d^2 \sigma _{pd}^{ - 2} \gamma _p^{ - 1} \Lambda ^{ - 1} )^{|{\cal{D}}_n |} }}e^{1/(\sigma _{pd}^2 \gamma _p )},
\end{split}\label{C.5}\tag{C.5}
\end{equation}
where the second equality is obtained by using $\frac{x}{{\sigma _d^2 }} + \frac{x}{{\sigma _{pd}^2 \gamma _p \Lambda }} = t$. Hence, we have completed the proof of (53) and (54) as (C.3) and (C.5), respectively.

\begin{IEEEbiography}[{\includegraphics[width=1in,height=1.25in]{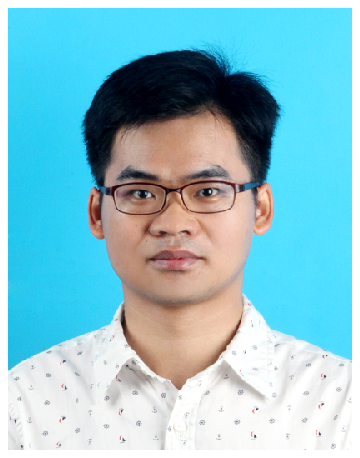}}]{Yulong Zou} (S'07-M'12-SM'13) is a Full Professor at the Nanjing University of Posts and Telecommunications (NUPT), Nanjing, China. He received the B.Eng. degree in Information Engineering from NUPT, Nanjing, China, in July 2006, the first Ph.D. degree in Electrical Engineering from the Stevens Institute of Technology, New Jersey, the United States, in May 2012, and the second Ph.D. degree in Signal and Information Processing from NUPT, Nanjing, China, in July 2012. His research interests span a wide range of topics in wireless communications and signal processing, including the cooperative communications, cognitive radio, wireless security, and energy-efficient communications.

Dr. Zou is currently serving as an editor for the IEEE Communications Surveys \& Tutorials, IEEE Communications Letters, EURASIP Journal on Advances in Signal Processing, and KSII Transactions on Internet and Information Systems. He served as the lead guest editor for a special issue on ``Security Challenges and Issues in Cognitive Radio Networks" in the EURASIP Journal on Advances in Signal Processing. He is also serving as the lead guest editor for a special issue on ``Security and Reliability Challenges in Industrial Wireless Sensor Networks" in the IEEE Transactions on Industrial Informatics. In addition, he has acted as symposium chairs, session chairs, and TPC members for a number of IEEE sponsored conferences, including the IEEE Wireless Communications and Networking Conference (WCNC), IEEE Global Communications Conference (GLOBECOM), IEEE International Conference on Communications (ICC), IEEE Vehicular Technology Conference (VTC), International Conference on Communications in China (ICCC), and so on.

\end{IEEEbiography}

\begin{IEEEbiography}[{\includegraphics[width=1in,height=1.25in]{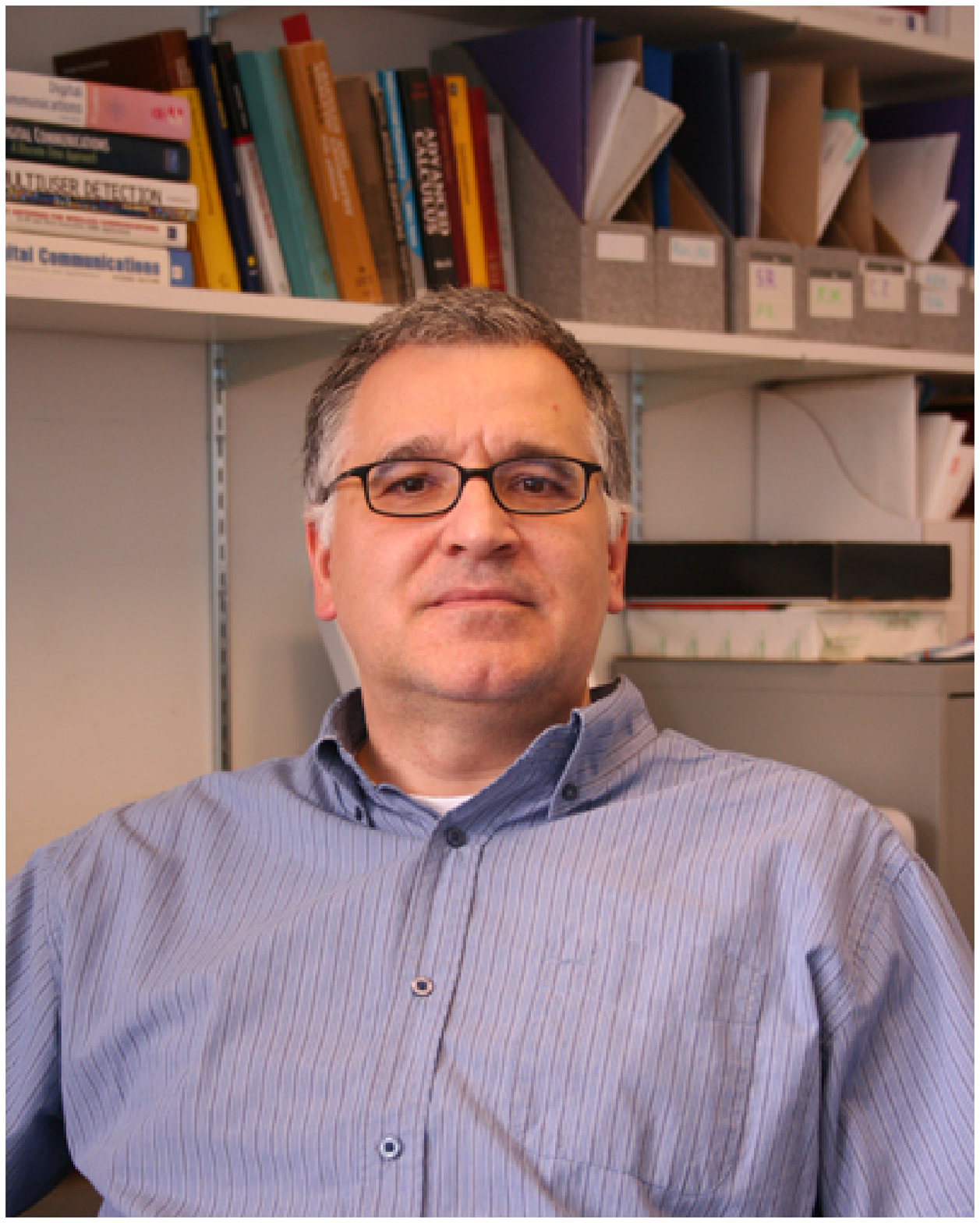}}]{Benoit Champagne}(S'87-M'89-SM'03) was born in Joliette (PQ), Canada, in 1961. He received the B.Ing. Degree in Engineering Physics and the M.Sc. Degree in Physics from the University of Montreal in 1983 and 1985, respectively, and the Ph.D. Degree in Electrical Engineering from the University of Toronto in 1990. From 1990 to 1999, he was with INRS, University of Quebec, where he held the positions of Assistant and then Associate Professor. In 1999, he joined McGill University, Montreal, as an Associate Professor with the Department of Electrical and Computer Engineering. He served as Associate Chairman of Graduate Studies in the Department from 2004 to 2007 and is now a Full Professor.

His research interests focus on the investigation of new computational algorithms for the digital processing of information bearing signals and overlap many sub-areas of statistical signal processing, including: detection and estimation, sensor array processing, adaptive filtering, multirate systems, and applications thereof to broadband voice and data communications. Over the years, he has supervised many graduate students in these areas and co-authored several papers, including key works on subspace tracking, speech enhancement, time delay estimation and spread sources localization.

\end{IEEEbiography}

\begin{IEEEbiography}[{\includegraphics[width=1in,height=1.25in]{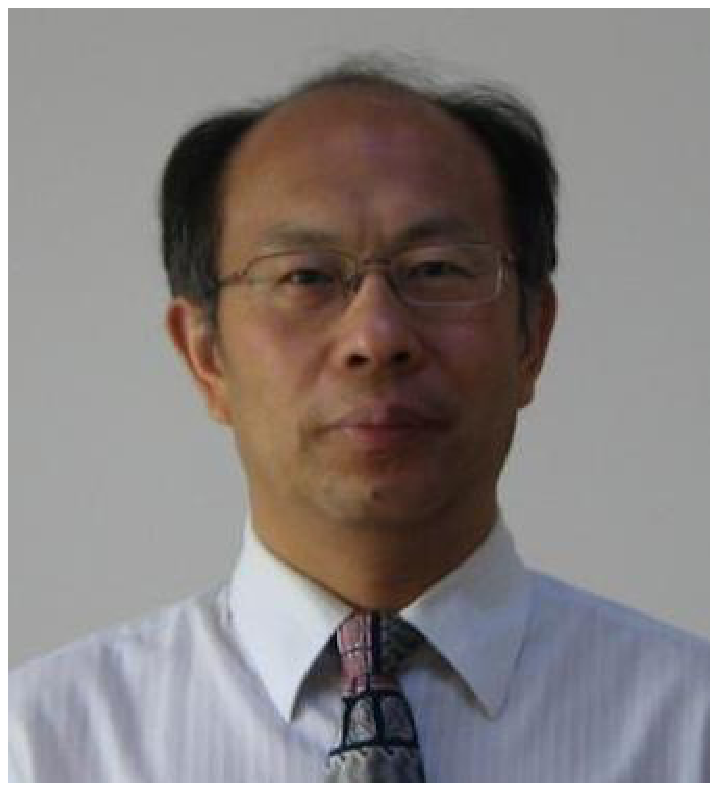}}]{Wei-Ping Zhu} (SM'97) received the B.E. and M.E. degrees from Nanjing University of Posts and Telecommunications, and the Ph.D. degree from Southeast University, Nanjing, China, in 1982, 1985, and 1991, respectively, all in electrical engineering. He was a Postdoctoral Fellow from 1991 to 1992 and a Research Associate from 1996 to 1998 with the Department of Electrical and Computer Engineering, Concordia University, Montreal, Canada. During 1993-1996, he was an Associate Professor with the Department of Information Engineering, Nanjing University of Posts and Telecommunications. From 1998 to 2001, he worked with hi-tech companies in Ottawa, Canada, including Nortel Networks and SR Telecom Inc. Since July 2001, he has been with Concordia's Electrical and Computer Engineering Department as a full-time faculty member, where he is presently a Full Professor. His research interests include digital signal processing fundamentals, speech and audio processing, and signal processing for wireless communication with a particular focus on MIMO systems and cooperative relay networks.

Dr. Zhu was an Associate Editor of the IEEE Transactions on Circuits and Systems Part I: Fundamental Theory and Applications from 2001 to 2003, and an Associate Editor of Circuits, Systems and Signal Processing from 2006 to 2009. He was also a Guest Editor for the IEEE Journal on Selected Areas in Communications for the special issues of: Broadband Wireless Communications for High Speed Vehicles, and Virtual MIMO during 2011-2013. Since 2011, he has served as an Associate Editor for the IEEE Transactions on Circuits and Systems Part II: Express Briefs. Dr. Zhu was the Secretary of Digital Signal Processing Technical Committee (DSPTC) of the IEEE Circuits and System Society during 2012-2014, where he is presently the Chair of the DSPTC.

\end{IEEEbiography}

\begin{IEEEbiography}[{\includegraphics[width=1in,height=1.25in]{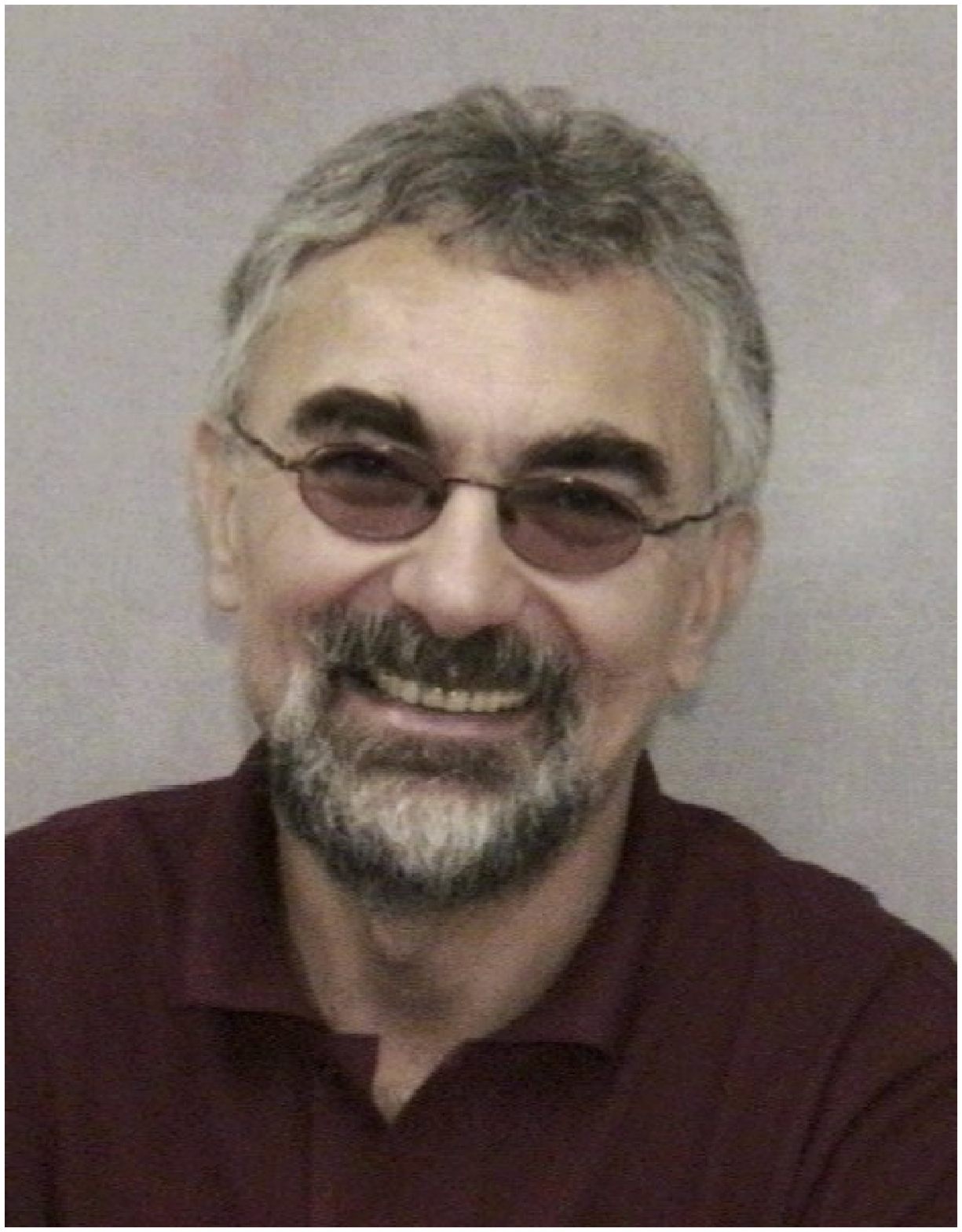}}]{Lajos Hanzo} (http://www-mobile.ecs.soton.ac.uk) FREng, FIEEE, FIET, Fellow of EURASIP, DSc received his degree in electronics in 1976 and his doctorate in 1983.  In 2009 he was awarded the honorary doctorate ``Doctor Honoris Causa'' by the Technical University of Budapest.  During his 37-year career in telecommunications he has held various research and academic posts in Hungary, Germany and the UK. Since 1986 he has been with the School of Electronics and Computer Science, University of Southampton, UK, where he holds the chair in telecommunications.  He has successfully supervised 80+ PhD students, co-authored 20 John Wiley/IEEE Press books on mobile radio communications totalling in excess of 10 000 pages, published 1400+ research entries at IEEE Xplore, acted both as TPC and General Chair of IEEE conferences, presented keynote lectures and has been awarded a number of distinctions. Currently he is directing a 100-strong academic research team, working on a range of research projects in the field of wireless multimedia communications sponsored by industry, the Engineering and Physical Sciences Research Council (EPSRC) UK, the European Research Council's Advanced Fellow Grant and the Royal Society's Wolfson Research Merit Award.  He is an enthusiastic supporter of industrial and academic liaison and he offers a range of industrial courses.  He is also a Governor of the IEEE VTS.  During 2008 - 2012 he was the Editor-in-Chief of the IEEE Press and a Chaired Professor also at Tsinghua University, Beijing.  His research is funded by the European Research Council's Senior Research Fellow Grant.  For further information on research in progress and associated
publications please refer to http://www-mobile.ecs.soton.ac.uk Lajos has 20 000+ citations.
\end{IEEEbiography}


\begin{thebibliography}{11}

\footnotesize

\bibitem{IEEEhowto:1}
J. Mitola and G. Q. Maguire, \textquotedblleft Cognitive radio: Making software radios more personal," \emph{IEEE Personal Commun.}, vol. 6, no. 4, pp. 13-18, Aug. 1999.

\bibitem{IEEEhowto:2}
IEEE 802.22 Working Group, ``IEEE P802.22/D1.0 draft standard for wireless regional area networks part 22: cognitive wireless RAN medium access control (MAC) and physical layer (PHY) specifications: policies and procedures for operation in the TV bands," Apr. 2008.

\bibitem{IEEEhowto:3}
G. Baldini, T. Sturman, A. R. Biswas, and R. Leschhorn, ``Security aspects in software defined radio and cognitive radio networks: A survey and a way ahead," \emph{IEEE Commun. Surv. \& Tut.}, vol. 14, no. 2, pp. 355-379, May 2012.

\bibitem{IEEEhowto:4}
D Cabric, S.M. Mishra, and R. W. Brodersen, ``Implementation issues in spectrum sensing for cognitive radios," in \emph{Proc. 2004 38th Asil. Conf. Sig., Sys. \& Comp.}, Pacific Grove, CA, Nov. 2004, pp. 772-776.

\bibitem{IEEEhowto:5}
H. Li, ``Cooperative spectrum sensing via belief propagation in spectrum-heterogeneous cognitive radio systems," in \emph{Proc. 2010 IEEE Wireless Commun. and Netw. Conf. (WCNC)}, Sydney Australia, Apr. 2010.

\bibitem{IEEEhowto:6}
J. Ma, G. Zhao, and Y. Li, ``Soft combination and detection for cooperative spectrum sensing in cognitive radio networks," \emph{IEEE Trans. Wireless Commun.}, vol. 7, no. 11, pp. 4502-4507, Nov. 2008.

\bibitem{IEEEhowto:7}
A. Ghasemi and E.S. Sousa, ``Fundamental limits of spectrum-sharing in fading environments," \emph{IEEE Trans Wireless Commun.}, vol. 6, no. 2, pp. 649-658, Feb. 2007.

\bibitem{IEEEhowto:8}
R. Southwell, J. Huang, and X. Liu, ``Spectrum mobility games," in \emph{Proc. 31\textsuperscript{st} Conf. Comp. Commun. (INFOCOM)}, Orlando, FL, Mar. 2012, pp. 37-45.

\bibitem{IEEEhowto:9}
I. F. Akyildiz, W.-Y. Lee, M. C. Vuran, and S. Mohanty, ``A survey on spectrum management in cognitive radio networks," \emph{IEEE Commun. Mag.}, vol. 46, no. 4, pp. 40-48 , Apr. 2008.

\bibitem{IEEEhowto:10}
H. Li and Z. Han, ``Dogfight in spectrum: Combating primary user emulation attacks in cognitive radio systems, part I: Known channel statistics," \emph{IEEE Trans. Wireless Commun.}, vol. 9, no. 11, pp. 3566-3577, Nov. 2010.

\bibitem{IEEEhowto:11}
T. Brown and A. Sethi, \textquotedblleft Potential cognitive radio denial-of-service vulnerabilities and protection countermeasures: a multi-dimensional analysis and assessment," in \emph{Proc. 2007 2nd Intern. Conf. Cogn. Radio Orient. Wirel. Net. and Commun. (CROWNCOM 2007)}, Orlando, FL, Aug. 2007, pp. 456-464.

\bibitem{IEEEhowto:12}
S. Lakshmanan, C. Tsao, R. Sivakumar, and K. Sundaresan, ``Securing wireless data networks against eavesdropping using smart antennas," in \emph{Proc. 28\textsuperscript{th} Intern. Conf. Dist. Comput. Sys. (ICDCS)}, Beijing, China, Jun. 2008, pp. 19-27.

\bibitem{IEEEhowto:13}
A. Olteanu and Y. Xiao, ``Security overhead and performance for aggregation with fragment retransmission (AFR) in very high-speed wireless 802.11 LANs," \emph{IEEE Trans. Wireless Commun.}, vol. 9, no. 1, pp. 218-226, Jan. 2010.

\bibitem{IEEEhowto:14}
Y. Xiao, V. K. Rayi, X. Du, F. Hu, and M. Galloway, ``A survey of key management schemes in wireless sensor networks," \emph{Comp. Commun.}, vol. 30, no. 11-12, pp. 2314-2341, Sept. 2007.

\bibitem{IEEEhowto:15}
A. Mukherjee, S. A. Fakoorian, J. Huang, and A. L. Swindlehurst, ``Principles of physical layer security in multiuser wireless networks: A survey," \emph{IEEE Commun. Surv. \& Tutor.}, vol. 16, no. 3, pp. 1550-1573, Aug. 2014.

\bibitem{IEEEhowto:16}
A. D.Wyner, ``The wire-tap channel," \emph{Bell System Technical Journal}, vol. 54, no. 8, pp. 1355-1387, 1975.

\bibitem{IEEEhowto:17}
S. K. Leung-Yan-Cheong and M. E. Hellman, ``The Gaussian wiretap
channel," \emph{IEEE Trans. Inf. Theory}, vol. 24, no. 4, pp. 451-456, Jul. 1978.

\bibitem{IEEEhowto:18}
P. Parada and R. Blahut, ``Secrecy capacity of SIMO and slow fading channels,"  in \emph{Proc. 2005 IEEE Intern. Sympos. Inform. Theory (IEEE ISIT 2005)}, Adelaide, SA, Sept. 2005, pp. 2152-2155.

\bibitem{IEEEhowto:19}
M. Bloch, J. O. Barros, M. R. D. Rodrigues, and S. W. McLaughlin, ``Wireless information-theoretic security," \emph{IEEE Trans. Inf. Theory}, vol. 54, no. 6, pp. 2515-2534, Jun. 2008.

\bibitem{IEEEhowto:20}
P. K. Gopala, L. Lai, and H. Gamal, ``On the secrecy capacity of fading channels ," \emph{IEEE Trans. Inf. Theory}, vol. 54, no. 10, pp. 4687-4698, Oct. 2008.

\bibitem{IEEEhowto:21}
Z. Li, W. Trappe, and R. Yates, ``Secret communication via multi-antenna transmission," in \emph{Proc. 41st Conf. Information
Sciences Systems}, Baltimore, MD, Mar. 2007, pp. 905-910.

\bibitem{IEEEhowto:22}
F. Oggier and B. Hassibi, ``The secrecy capacity of the MIMO wiretap channel," \emph{IEEE Trans. Inf. Theory}, vol. 57, no. 8, pp. 4961-4972, Oct. 2007.

\bibitem{IEEEhowto:23}
M. Yuksel and E. Erkip, ``Secure communication with a relay helping the wiretapper," in \emph{Proc. 2007 IEEE Information Theory Workshop}, Lake Tahoe, CA, Sep. 2007, pp. 595-600.

\bibitem{IEEEhowto:24}
{L. Dong, Z. Han, A. P. Petropulu, and H. V. Poor, ``Improving wireless physical layer security via cooperating relays," \emph{IEEE Trans. Signal Process.}, vol. 58, no. 3, pp. 1875-1888, Mar. 2010.}

\bibitem{IEEEhowto:25}
{Y. Zou, X. Wang, and W. Shen, ``Optimal relay selection for physical-layer security in cooperative wireless networks," \emph{IEEE J. Sel. Areas Commun.}, vol. 31, no. 10, pp. 2099-2111, Oct. 2013.}

\bibitem{IEEEhowto:26}
A. Mukherjee amd A. Swindlehurst, ``Robust beamforming for security
in MIMO wiretap channels with imperfect CSI," \emph{IEEE
Trans. Signal Process.}, vol. 59, no. 1, pp. 351-361, Jan. 2011.

\bibitem{IEEEhowto:27}
C. Jeong, I. Kim, and K. Dong, ``Joint secure beamforming design at
the source and the relay for an amplify-and-forward MIMO untrusted
relay system," \emph{IEEE Trans. Signal Process.}, vol. 60, no. 1, pp. 310-325, Jan. 2012.

\bibitem{IEEEhowto:28}
Y. Pei, Y.-C. Liang, K.C. Teh, and K. Li, ``Secure communication in multiantenna cognitive radio networks with imperfect channel state information," \emph{IEEE Trans. Signal Process.}, vol. 59, no. 4, pp. 1683-1693, Apr. 2011.

\bibitem{IEEEhowto:29}
Y. Zou, X. Wang, and W. Shen, ``Physical-layer security with multiuser scheduling in cognitive radio networks", \emph{IEEE Trans. Commu.}, vol. 61, no. 12, pp. 5103-5113, Dec. 2013.

\bibitem{IEEEhowto:30}
{Z. Shu, Y. Qian, and S. Ci, ``On physical layer security for cognitive radio networks," \emph{IEEE Net. Mag.}, vol. 27, no. 3, pp. 28-33, Jun. 2013.}

\bibitem{IEEEhowto:31}
Y. Zou, X. Wang, W. Shen, and L. Hanzo, ``Security versus reliability analysis of opportunistic relaying," \emph{IEEE Trans. Veh. Tech.}, vol. 63, no. 6, pp. 2653-2661, Jun. 2014.

\bibitem{IEEEhowto:32}
L. Di Stefano and S. Mattoccia, ``A sufficient condition based on the Cauchy-Schwarz inequality for efficient template matching," \emph{Proc. 2003 Intern. Conf. Image Process.}, Catalonia, Spain, Sept. 2003, pp. 269-272.

\bibitem{IEEEhowto:33}
M. Abramowitz and I. A. Stegun. \emph{Handbook of Mathematical Functions with Formulas, Graphs, and Mathematical Tables}, Ninth
Edition, New York: Dover Publications, 1970.

\bibitem{IEEEhowto:34}
{Y. Zou, Y.-D. Yao, and B. Zheng, ``Diversity-multiplexing tradeoff in selective cooperation for cognitive radio," \emph{IEEE Trans. Commun.}, vol. 60, no. 9, pp. 2467-2481, Sept. 2012.}

\bibitem{IEEEhowto:35}
{S. Goel and R. Negi, ``Guaranteeing secrecy using artificial noise," \emph{IEEE Trans. Wireless Commun.}, vol. 7, no. 6, pp. 2180-2189, Jul. 2008.}

\bibitem{IEEEhowto:36}
{W. Li, M. Ghogho, B. Chen, and C. Xiong, ``Artificial noise by the receiver: Outage secrecy capacity/region analysis," \emph{IEEE Commun. Lett.}, vol. 16, no. 10, pp. 1628-1631, Oct. 2012.}

\end{thebibliography}
\end{document}